\documentclass[rmp,aps,superscriptaddress,nofootinbib,twocolumn]{revtex4-1}

\usepackage{amsmath}
\usepackage{graphicx}
\usepackage{color}

\begin{document}

\title{Colloquium: Cluster growth on surfaces - densities,
size distributions and morphologies}

\author{Mario Einax}
\email{mario.einax@uni-osnabrueck.de}
\affiliation{Fachbereich Physik, Universit\"at Osnabr\"uck,
Barbarastra{\ss}e 7, 49076 Osnabr\"uck, Germany}
\author{Wolfgang Dieterich}
\email{wolfgang.dieterich@uni-konstanz.de}
\affiliation{Fachbereich Physik, Universit\"at Konstanz, 78457
Konstanz, Germany}
\author{Philipp Maass}
\email{philipp.maass@uni-osnabrueck.de}
\affiliation{Fachbereich Physik, Universit\"at Osnabr\"uck,
Barbarastra{\ss}e 7, 49076 Osnabr\"uck, Germany}

\date{\today}

\begin{abstract}
  Understanding and control of cluster and thin film growth on solid
  surfaces is a subject of intensive research to develop nanomaterials
  with new physical properties. In this Colloquium we review basic
  theoretical concepts to describe submonolayer growth kinetics under
  non-equilibrium conditions. It is shown how these concepts can be
  extended and further developed to treat self-organized cluster
  formation in material systems of current interest, such as
  nanoalloys and molecular clusters in organic thin film growth. The
  presentation is focused on ideal flat surfaces to limit the scope
  and to discuss key ideas in a transparent way. Open experimental and
  theoretical challenges are pointed out.
\end{abstract}

\maketitle
\tableofcontents

\section{Introduction}
\label{Introduction}
A widely applied method for the design of materials with nanoscale
dimensions is to deposit atoms or molecules on a solid substrate.
Adsorbed atoms diffuse along the surface, nucleate and form
aggregates. The interplay of these processes leads to a rich variety
of self-organized growth phenomena. Nanoscale structures built in this
way in general are \emph{metastable}, so that their physical
properties can differ distinctly from the corresponding equilibrium
bulk phases. Owing to the progress in experimental techniques,
especially scanning tunneling microscopy (STM) and atomic force
microscopy (AFM), it is possible to uncover microscopic details in the
underlying structure formation with an unprecedented precision.

Starting out with the pioneering work by Venables and
coworkers~\cite{Venables:1973, Venables/etal:1984} important
theoretical concepts were developed in the past. Today, the adatom
kinetics of single-component metallic systems is rather well
understood. From scaling properties of measured island densities and
shapes important kinetic parameters can be extracted, like adatom
diffusion coefficients and interaction energies, sizes of critical
nuclei, step edge barriers for interlayer transport, etc. Knowledge of
these parameters makes it possible to control the desired growth modes
to a substantial degree. Several extensive review articles give an
excellent account of the state-of-the-art~\cite{Brune:1998,
  Ratsch/Venables:2003, Michely/Krug:2004, Evans/etal:2006}.

This Colloquium is motivated by the fact that many of the basic
questions in the field of surface growth appear nowadays in a new
context. This includes self-organized growth of \emph{nanoalloys},
which can show unexpected physical properties emerging from frozen-in
non-equilibrium atomic arrangements. The submonolayer regime of alloy
growth offers a wealth of new problems relative to growth in
one-component adatom systems. Other examples of wide current interest
are epitaxial growth of colloids \cite{Ganapathy/etal:2010,
  Einstein/Stasevich:2010}, growth of graphene on metal substrates
\cite{Zangwill/Vvedensky:2011, Loginova/etal:2008,Coraux/etal:2009},
and growth of metallic nanoparticles on graphene
\cite{Zhou/etal:2010,Pandey/etal:2011,Moldovan/etal:2012}. While
theoretical treatments of the growth kinetics for these systems are
still rare, the general concepts, presented here, can be adapted and
refined in order to understand the graphene and nanoparticle
formation.

A further important problem is the self-organized growth of
\emph{organic molecules} on an inorganic
substrate~\cite{Kowarik/etal:2008, Kuehnle:2009,Rahe/etal:2013}. In
the latter case, the complexity of building blocks, provided by
organic chemistry, can be exploited for generating an even richer
spectrum of surface structures. Such studies are largely driven by
perspectives of molecular electronics \cite{Cuevas/Scheer:2010,
  Cuniberti/etal:2005, Nitzan/Ratner:2003} and organic photovoltaics
\cite{Deibel/Dyakonov:2010, Nicholson/Castro:2010}.

A challenge in this area is to clarify how far concepts based on
single atom surface kinetics remain valid, or require modification
when dealing with the larger sizes, potential non-spherical shapes and
internal degrees of freedom of the molecules. Such questions were
addressed in recent experiments, but our understanding of mechanisms
underlying organic surface growth is just at the beginning. Of
particular relevance is to explain which type of island morphologies
develop and how they can be controlled. Organic molecules exhibit
often only weak interactions with the substrate. As a consequence,
dewetting is often observed in molecular pattern formation
\cite{Burke/etal:2009}). Examples are pentacene on SiO$_2$
\cite{Kaefer/etal:2009}, perylene tetracarboxylic dianhydride (PTCDA)
on NaCl \cite{Burke/etal:2008}, or C$_{60}$ on KBr
\cite{Burke/etal:2005}, NaCl \cite{Burke/etal:2007} and on CaF$_2$
\cite{Loske/etal:2010}. Finally, it is fair to note that even the
standard theories of cluster growth in mono-component atomic systems
still imply basic, partly long-standing open problems and we will
point out a few of them.

\begin{figure}[t!]
\includegraphics[width=0.45\textwidth,clip=,]{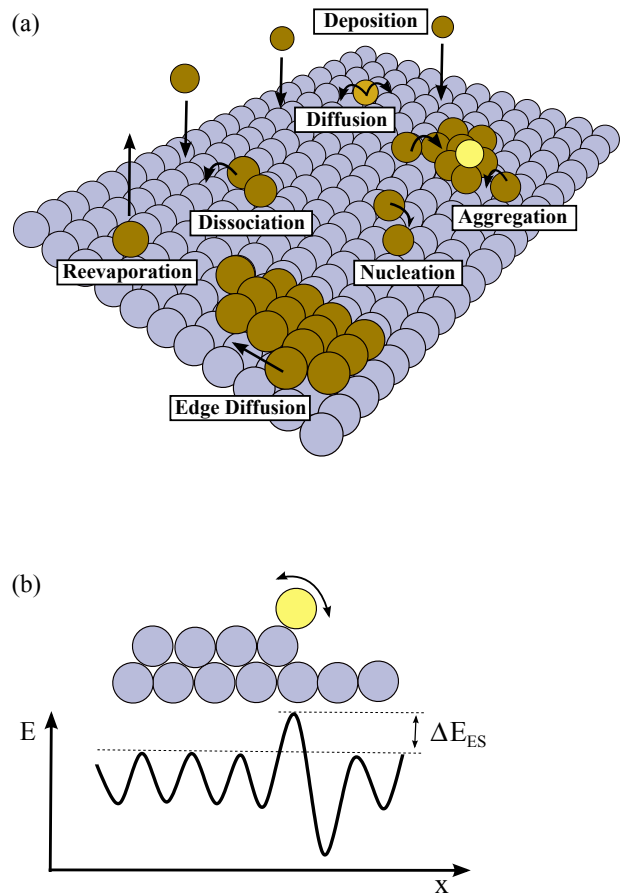}
\caption{(Color online) Illustration of (a) elementary moves during
  growth after deposition onto a surface and (b) the Ehrlich-Schwoebel
  barrier for crossing the step edge.}
\label{fig:fig1}
\end{figure}

Our focus will be on a set of general questions, which seemingly are
well understood but indeed call for a more general treatment or deeper
analysis: (i) How many islands form? (ii) What is the distribution of
island sizes? (iii) When do islands survive upon continued deposition?
(iv) What determines the outer shape and inner structure of clusters?
The first two questions are specially important for submonolayer
growth and the following ones for multilayer growth. Without reviewing
in any detail the current knowledge about these questions, our
attention in each case will be concentrated on specific open problems
which are behind. Steps towards their solution are proposed,
corroborated by simple model studies.

\section{Elementary processes and rate equations}
\label{sec:rate-equations}
Advanced computer simulation techniques have made it possible to study
the growth of clusters and films on surfaces in great detail [for a
recent review, see \textcite{Clancy:2011}]. Classical methods as
molecular dynamics \cite{Rapaport:2005} and kinetic Monte-Carlo (KMC)
simulations \cite{Landau/Binder:2005} can nowadays be supported by
\textit{ab initio} calculations, which provide precise force fields
and/or microscopic parameters entering rates for elementary jump
processes. Such approaches \cite{Kratzer/Scheffler:2001} are important
to account for specifics of material systems. To capture large time
and length scales in the growth kinetics, continuum phase-field models
have been developed, where atomic features are resolved up to a
certain extent \cite{Provatas/Elder:2010}. Prominent approaches are
the level set method \cite{Gyure/etal:1998,Ratsch/Venables:2003} and
the phase-field crystal method
\cite{Elder/etal:2002,Greenwood/etal:2010}. The latter is based on a
modulated density field that minimizes a free energy functional.
Construction of this functional to adequately describe patterning in
ultra thin films is the major challenge of a phase-field crystal
method \cite{Elder/etal:2012}.

To bring up the basic principles in a transparent way, we consider
ideal surfaces, that is we disregard all complicating factors arising
at real surfaces, as, for example, surface steps, impurities,
anisotropies or reconstruction. Given this scope, surface structures
evolve through elementary atomic or molecular moves that obey fairly
simple rules, but a sequence of a large number of moves eventually
leads to a high degree of self-organization and to complex terminal
structures on the nanoscopic or even mesoscopic scale.

Figure~\ref{fig:fig1}(a) illustrates some types of atomic moves, which
dominate the early stages of growth. Following deposition to the
substrate surface, atoms can reevaporate, or stick to the surface and
perform diffusional steps. Diffusing adatoms (monomers) can stick
together when they meet, forming dimers, trimers or larger
two-dimensional islands. Islands as a whole in general do not diffuse;
they grow by attachment of other adatoms or decay by dissociation.
Three-dimensional growth is due to direct deposition on top of an
island that has formed before, or to interlayer jumps. In an
interlayer jump, illustrated in Fig.~\ref{fig:fig1}(b), the atom in
general must surmount an additional energy barrier, the so-called
Ehrlich-Schwoebel barrier associated with a low-coordinated site at a
step edge \cite{Ehrlich/Hudda:1966, Schwoebel:1969}.

Structure formation begins with the assembly of two-dimensional
islands, composed of adsorbate atoms within the first monolayer.
Usually a range of fairly low temperatures exists, where reevaporation
can be ignored and thermal energies $k_{\rm B}T$ are significantly
lower than the binding energy $E_{\rm B}$ between two adatoms, while
diffusion is active. The condition $k_{\rm B}T \ll E_{\rm B}$ ensures
that only few atoms are required to form a stable island. An important
concept is the critical island size (\emph{critical nucleus}). Islands
composed of more than $i$ atoms are more likely to grow than to decay.

Two factors influence the nucleation and growth of islands:
deposition of atoms onto the surface with a flux $F$ and
thermally activated diffusion of adatoms along the surface with a
diffusion coefficient
\begin{equation}
\label{eq:Arrhenius}
D \simeq  D_{\infty} \exp(-U/k_{\rm B}T)\, .
\end{equation}
where $U$ is the diffusion barrier. For atomic systems, the
pre-exponential factor is given by $D_{\infty} =\nu a^2$, where $\nu$
is an attempt frequency and $a$ is the lattice constant of the
substrate. The mean time for a unit cell to be hit by an atom is
$1/Fa^2$, and $a^2/D$ is the mean time after which it leaves that cell
by diffusion. Growth kinetics are controlled by the dimensionless
ratio of these times, the ``$D/F$-ratio'' $\Gamma = D/Fa^4$.  In the
following, $a$ is used as length unit and not always given explicitly.

\begin{figure}[t!]
\includegraphics[width=0.32\textwidth,clip=,]{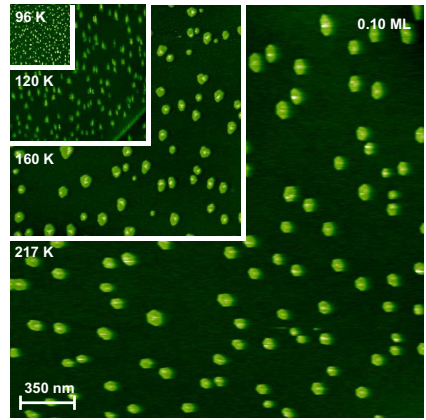}
\caption{(Color online) AFM images of C$_{60}$ molecules on
  CaF$_2$(111) at a coverage $\Theta=0.1$ for different substrate
  temperatures between $96$ and $217$~K.
  From~\textcite{Loske/etal:2010}.}
\label{fig:fig2}
\end{figure}

Because adatom diffusion is thermally activated, the density $N$ of
stable islands is a sensitive function of temperature. At high
temperatures, $D$ is large and adatoms can diffuse over longer
distances before encountering another adatom or attaching to an
island. Accordingly, $N$ becomes smaller with increasing $T$. This is
demonstrated in Fig.~\ref{fig:fig2} for fullerene (C$_{60}$) islands
grown on an atomically flat CaF$_2$(111) surface
\cite{Loske/etal:2010}.

Rate equations (RE) have proven to be very useful to relate measured
island densities to atomistic
parameters~\cite{Zinsmeister:1966,Venables:1973}. The approach follows
the spirit of classical nucleation theory by
\textcite{Becker/Doering:1935} and is based on equations for the
evolution of densities $n_s(t)$ of islands composed of $s$ atoms.
Monomers have a density $n_1(t) \equiv n(t)$. Islands with $s \ge 2$
are considered to be immobile. This leads to the infinite set of
equations
\begin{subequations}
\begin{align}
\label{eq:monomer_density}
\frac{d n}{d t} &=  F  - 2D \sigma_1 n^2 -
D n \sum_{s=2}^{\infty} \sigma_s n_s + 2 K_2 n_2 +
\sum_{s=3}^{\infty}  K_s  n_s\,,\\
\label{eq:stable_islands}
\frac{d n_s}{d t} &= D\sigma_{s-1}n n_{s-1}-D\sigma_{s}n n_{s}\nonumber\\
&\phantom{=}{}+  K_{s+1}  n_{s+1} - K_s  n_s\, ,
\hspace{1em} s=2,3,\ldots
\end{align}
\label{eq:rate1}\hspace*{-0.4em}
\end{subequations}
Attachment of diffusing monomers to $s$-clusters occurs with rates
$\sigma_s D n n_s$, where $\sigma_s$ are called capture numbers.
Detachment of monomers from $s$-clusters occurs with rates $K_s$,
where $K_s=0$ if $s>i$. Re-evaporation of atoms, direct impingement of
arriving atoms onto clusters, coalescence of clusters or dissociation
of clusters into subclusters are neglected in Eqs.~(\ref{eq:rate1})
but can be incorporated by proper
extension~\cite{Venables/etal:1984,Brune:1998,Venables:2000}.

A straightforward step in refining Eqs.~(\ref{eq:rate1}) is to
distinguish the impingement of atoms to the free surface from
impingement to the islands edge, or to the top of an island and
subsequent attachment to its edge. $F$ is then replaced by
$F(1-\Theta)$ and terms of the form $F\kappa_s n_s$ are added, where
$\kappa_s$ is a direct capture area \cite{Venables:1973}. In
Sec.~\ref{sec:size_distribution} we will refer to the corresponding
rate equations as the refined RE.

From the above concepts it is clear that cluster growth on surfaces
basically differs from growth in solutions. There, small clusters
composed of only a few atoms (molecules) are generally unstable due to
solvation effects. This implies that the critical nuclei are large, in
contrast to metallic surface growth. In standard continuum theory they
are characterized by a critical radius, which results from the
competition of the surface free energy, increasing as $R^2$, and the
bulk free energy, proportional to $R^{3}$. Cluster diffusion and
cluster aggregation that both are affected by hydrodynamic effects,
play an important role in distinction to processes shown in
Fig.~\ref{fig:fig1}. Moreover, crystal growth in solution involves
many metastable intermediates. The large variety of intermediate
structures generally has no analogue in surface growth. Finally, the
build-up of crystalline structures is associated with release of
latent heat, which requires the introduction of a temperature field
and heat diffusion equation into the theory.

\section{How many islands form?}
\label{sec:islands}

\subsection{Island densities of one-component adsorbates}
\label{subsec:one-component}
To predict from Eqs.~(\ref{eq:rate1}) the dependence of density
$N=\sum_{s=i+1}n_s$ of stable islands on $\Gamma$, it is sufficient to
replace the $\sigma_s$ for $s>i$ by an averaged capture number
$\bar{\sigma}$,
\begin{align}
\label{eq:averaged_capture}
\bar{\sigma}&= \frac{1}{N} \sum_{s=i+1}^{\infty} \sigma_s n_s \,.
\end{align}
In addition, a quasi-stationary state for unstable islands of size
$2\le s\le i$ is assumed, where decay and aggregation processes nearly
balance each other, $K_s n_s\simeq D \sigma_{s-1} n n_{s-1}$, yielding
$n_s\propto n^s$. Since a monomer, when it gets detached from an
$s$-cluster, has to overcome an energy barrier that is composed of
both its binding energy $(E_{s-1}-E_s)$ to the cluster and the
diffusion barrier, one can write $K_s=\mu_s
D\exp[-(E_s-E_{s-1})/k_{\rm{B}} T]$, where $\mu_s$ is a constant.
Here $E_s>0$ is the total energy needed to decompose an $s$-cluster
into monomers ($E_1=0$). In this way one arrives at the so-called
Walton relations \cite{Walton:1962}
\begin{align}
\label{eq:Walton_relation}
n_s&\simeq b_s\exp(E_s/k_{\rm{B}} T)\,n^s \,\hspace{1em}s=2,\ldots,i,
\end{align}
where $b_s=\prod_{j=1}^{s-1}\sigma_j/\mu_{j+1}$.

Summation of Eq.~(\ref{eq:stable_islands}) over $s>i$ leads to a
cancelation of all terms on the right hand side except the term $s=i$.
Using Eqs.~(\ref{eq:averaged_capture}) and (\ref{eq:Walton_relation})
the rate equations Eqs.~(\ref{eq:rate1}) reduce to a closed set of
equations for $n(t)$ and $N(t)$,
\begin{subequations}
\begin{align}
\label{eq:monomer_density_simplified}
\frac{d n}{d t} &=F-(1+\delta_{i,1})\sigma_{i} D n  n_{i} -
\bar{\sigma} D n N\,, \\
\label{eq:total_stable_islands}
\frac{d N}{d t} &= \sigma_{i} D n  n_{i} \, .
\end{align}
\label{eq:rate_simplified}\hspace*{-0.4em}
\end{subequations}
with $n_i\propto n^i$ from Eq.~(\ref{eq:Walton_relation}). The last
equation simply tells that $N(t)$ grows by nucleation events.

Experimental $\Gamma$ values are very large compared to unity,
typically $10^5< \Gamma < 10^{11}$. Values of that order are often
representative of the leading asymptotic behavior in the limit $\Gamma
\rightarrow \infty$. An asymptotic solution of
Eqs.~(\ref{eq:rate_simplified}) in this limit can be derived from the
scaling ansatz $n(\Gamma,\Theta)\sim \Gamma^{-\zeta}n_\infty(\Theta)$
and $N(\Gamma,\Theta)\sim \Gamma^{-\chi}N_\infty(\Theta)$ with
$\zeta,\chi>0$ \cite{Dieterich/etal:2008}.  Note that both $n$ and $N$
should decrease with increasing $\Gamma$ at fixed $\Theta$. Inserting
this scaling ansatz into Eqs.~(\ref{eq:rate_simplified}) gives
$\zeta=2/(i+2)$ and $\chi=i/(i+2)$. Finally,
\begin{subequations}
\begin{align}
N(\Gamma,\Theta)&\sim
\left[\frac{(i+2)\sigma_ib_i}{\bar\sigma^{i+1}}\,
    \Theta\right]^{\frac{\scriptstyle
    1}{\scriptstyle(i+2)}}e^{\frac{\scriptstyle E_i}{\scriptstyle (i+2)k_{\rm B}T}}
\Gamma^{-\frac{\scriptstyle i}{\scriptstyle
    (i+2)}}\,,\label{eq:N-scaling}\\
n(\Gamma,\Theta)&\sim\frac{1}{\bar\sigma\Gamma N(\Gamma,\Theta)}\,.
\label{eq:n-scaling}
\end{align}
\label{eq:nN-scaling}\hspace*{-0.4em}
\end{subequations}
Equation ~(\ref{eq:n-scaling}) is often used in the literature as
``quasi-stationary'' approximation in Eqs.~(\ref{eq:rate_simplified})
to derive Eq.~(\ref{eq:N-scaling}). Our derivation is more general and
shows that this quasi-stationary relation becomes exact in the
$\Gamma\to\infty$ limit. Its range of validity for finite $\Gamma$ can
be estimated from the first-order corrections $\Delta
N(\Gamma,\Theta)$ and $\Delta n(\Gamma,\Theta)$ to the solution
(\ref{eq:nN-scaling}). In the case $i=1$, for example, both ratios
$\Delta N(\Gamma, \Theta)/N(\Gamma, \Theta)$ and $\Delta n(\Gamma,
\Theta)/n(\Gamma, \Theta)$ are found to behave as $\sim (\Theta^2
\Gamma)^{-1/3}$, and the criterion for Eq.~(\ref{eq:N-scaling}) to
hold is
\begin{equation}
\label{eq:asymptotic_criterion}
(\Theta^2 \Gamma)^{1/3} \gg
\left(\frac{64\sigma_1}{9\bar\sigma^2}\right)^{1/3}\,,
\end{equation}
where the right hand side is of order unity. The smaller $\Theta$,
the larger $\Gamma$ in order for the scaling solutions
(\ref{eq:nN-scaling}) to remain valid.  For $i=1$ the ratio $N(\Gamma,
\Theta)/n(\Gamma, \Theta)$ also behaves as $\sim (\Theta^2
\Gamma)^{1/3}$ according to Eqs.~(\ref{eq:nN-scaling}). This means
that the applicability of Eqs.~(\ref{eq:nN-scaling}) is equivalent to
the experimentally testable condition $n(\Gamma,\Theta) \ll
N(\Gamma,\Theta)$. Because $\Gamma$ is large, this condition is well
satisfied in a large range of $\Theta$ before $N$ steeply falls in the
coalescence regime, which typically sets in near $\Theta \approx 0.2$.
In other words, the scaling of the island density with respect to
$\Gamma$ is not restricted to a ``saturation regime'' of nearly
$\Theta$-independent island density, although this regime preceding
island coalescence is certainly most convenient for experiments. An
alternative argument for obtaining the crossover to the stationary
regime has been given earlier by equating Eq.~(\ref{eq:N-scaling})
with the solution for $N$ in the initial time regime
\cite{Evans/etal:2006}.

\begin{figure}[t!]
\includegraphics[width=0.35\textwidth,clip=,]{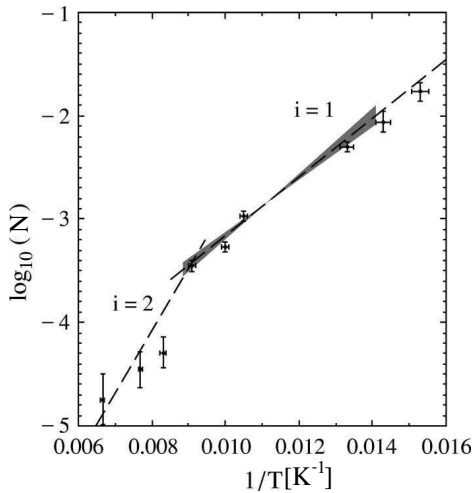}
\caption{Island densities of Ag on Pt$(111)$ at coverages
  $\Theta=0.12$ in an Arrhenius plot. Analysis of the slope with
  Eq.~(\ref{eq:N-scaling}) in the low-temperature regime, where $i=1$,
  gives a diffusion barrier $U =0.168\pm 0.005$~eV.  In the
  high-temperature regime, where $i=2$, the slope yields a dimer
  bonding energy $E_2=0.15\pm0.02$~eV.
  From~\textcite{Brune/etal:1999}.}
\label{fig:fig3}
\end{figure}

Measurements of $N$ as a function of $F$ and $T$ allow one to extract
critical island sizes, diffusion coefficients and binding energies.
Via the $\Gamma$-dependent factor in Eq.~(\ref{eq:N-scaling}) one
first determines $i$. In practice, because $T$ competes with the
binding energy $E_{\rm B}$, $i$ is constant only within certain
temperature intervals and overall increases with $T$. This is
illustrated in Fig.~\ref{fig:fig3} for the system
Ag/Pt(111)~\cite{Brune:1998}.  Knowing $i$, the slopes in an Arrhenius
representation of $N$ yield $(iU+E_i)/(i+2)$ as activation energy,
which for $i=1$ is $U/3$. The high-$T$ extrapolation of $N$ in
addition allows one to obtain information about the pre-exponential
factor $D_\infty$. This ``nucleation route'' \cite{Brune/etal:1994,
  Brune:1998, Barth/etal:2000, Mueller/etal:1996} to measuring surface
diffusion coefficients and binding energies has been found particular
useful. Examples are diffusion of Pd diffusion on SrTiO$_3$(100)
\cite{Richter/Wagner:2005}, of Co diffusion on Cu(111)
\cite{Prieto/etal:2000}, or of C$_{60}$ molecules on CaF$_2$(111)
\cite{Loske/etal:2010}, and of hydrogenated tetraphenyl porphyrin
(2H-PPT) on Ag(111) \cite{Rojas/etal:2011}. A thorough overview on
similar analyses of surface diffusion on metals can be found in
\textcite{Antczak/Ehrlich:2010}.

For a quantitative description of $N$ beyond $\Gamma$-scaling, it
turns out that the approximation (\ref{eq:averaged_capture}) of
constant capture numbers is insufficient. In particular the $\Theta$
dependence predicted by Eq.~(\ref{eq:N-scaling}) does not account for
the tendency for $N$ to saturate at large $\Theta$ before the
coalescence regime. A good quantitative account is provided by a
self-consistent theory of the capture numbers, which will be discussed
in Sec.~\ref{subsec:self-consistent} below.

The rate equations form a very useful basis for treating problems in
more complicated situations. Extensions of the theory for predicting
island densities of multi-component adsorbates are discussed in the
following Secs.~\ref{subsec:two-component} and
\ref{subsec:self-consistent}. A beautiful example is the growth of
graphene on metallic surfaces, induced by deposition of carbon atoms
or hydrocarbon molecules.  Nucleation can then proceed via a two-step
process. Clusters of five carbon atoms, which still are mobile, form
first. In turn, six such clusters react to a stable, immobile graphene
island~\cite{Zangwill/Vvedensky:2011}. Measured carbon adatom
densities on Ru(0001) \cite{Loginova/etal:2008} could be analyzed in
terms of coupled rate equations for the densities of adatoms ($n$),
five-atom clusters ($c$), and stable islands ($N$), thereby giving
insight into the detailed kinetics of this more complicated nucleation
scenario. Because of the large ``critical nuclei'' in this case, one
has $n \gg c \gg N$. Note that due to the scaling
$n/N\sim\Gamma^{(i-2)/(i+2)}$, cf.\ Eq.~(\ref{eq:nN-scaling}), $n\gg
N$ is in general expected to hold for $i>2$.

Epitaxial growth of graphene was indeed observed on several metallic
surfaces (\cite{Neto/etal:2009,Voloshina/Dedkov:2010}), such as
Ru(0001) \cite{Loginova/etal:2008,Zhou/etal:2010}, Ir(111)
\cite{Coraux/etal:2009,Rusponi/etal:2010}, Pt(111)
\cite{Sutter/etal:2009,Gao/etal:2011},
Ni(111)\cite{Grueneis/Vyalikh:2008}, and Cu(111)
\cite{Gao/etal:2010}). Single-layer graphene on metal surfaces as
well as chemically modified graphene and few-layer graphene can
conversely be used as substrate for growth of metals like Ag, Au, Fe,
Pt, and Ti. Such substrates modify the energetics of adsorbed metal
atoms, allowing the production of a variety of metallic nanoparticle
structures [for details, see \textcite{Pandey/etal:2011}, and
\textcite{Moldovan/etal:2012}].

\subsection{Island densities of binary alloys}
\label{subsec:two-component}
For binary (or multi-component) alloys, growth kinetics are controlled
by an enlarged set of parameters including mixing ratios and surface
diffusion coefficients, which can vary strongly among different atomic
species. This poses the question of a multi-parameter scaling of
island densities. Moreover, atoms of different type will differ in
their mutual binding energies on the surface, which can lead to a
competition of different nucleation paths. The coexistence of critical
nuclei with different compositions together with asymmetries in adatom
diffusion coefficients leads to new cross-over phenomena in the
scaling of island densities. Detection of such cross-overs by
experiment and interpretation with the help of theoretical
predictions, described below, should allow one to deduce novel
information on the size and composition of critical nuclei, in
particular on the binding energies of unlike atoms in the presence of
the surface.

Let us consider two species of atoms, $\alpha=A$ and $B$ to be
co-deposited with partial fluxes $F_\alpha=x_\alpha F$, where
$F=F_{\rm A}+F_{\rm B}$ is the total flux. In this way a binary alloy
with mole fractions $x_{\rm A}$ and $x_ {\rm B}=1-x_{\rm A}$ is
formed.  Only monomers of densities $n_\alpha$ are supposed to diffuse
along the surface with diffusion coefficients $D_\alpha = \nu_\alpha
\exp (-U_\alpha/k_{\rm B} T)$, while clusters with more than one atom
($s>1$) are immobile.

To elucidate the essential new physics compared to one-component
systems, we limit our discussion to cases where the largest unstable
clusters are composed of not more than two atoms (for the general
treatment, see~\textcite{Einax/etal:2007a}). Stable islands of
density $N$, in particular trimers and larger clusters, are treated in
an averaged manner irrespective of composition. Monomers with
densities $n_{\alpha}$ and all combinations of dimers (densities
$n_{\alpha \beta}$) will be treated explicitly. Considering systems
where the important differences between $A$ and $B$ atoms primarily
arise from diffusion coefficients and fluxes, we do not distinguish
between $A$ and $B$ atoms in their respective capture numbers.

In comparison with Eq.~(\ref{eq:monomer_density}) the equation of
motion for $A$ monomers involves additional terms due to capture of
$B$ atoms and to the decay of $AB$ dimers. Similarly the rate
equations for the dimer density involve additional terms. The
structure of these equations is quite obvious. For example, for
$n_{\rm AB}(t)$ the equation is
\begin{align}\label{eq:rate2}
  \frac{dn_{\rm AB}}{dt} &= (D_{\rm A} + D_{\rm B}) \sigma_1 n_{\rm A}
  n_{\rm B} \nonumber\\
&{}-\bigl(\sum_\beta
  D_\beta n_\beta\bigr) \sigma_2 n_{\rm AB} - K_{\rm AB}n_{\rm AB}.
\end{align}
Note that the coefficient for relative diffusional motion of $A$ and
$B$ monomers is $D_{\rm A}+D_{\rm B}$. $K_{\alpha
  \beta}=K_{\beta\alpha}$ denote the dissociation rates of dimers. A
reduced set of rate equations for $n_{\alpha}(t)$, $n_{\alpha
  \beta}(t)$ and $N(t)$ is thus obtained, which allows us to derive
the scaling behavior of island densities.

In experiments with binary systems both conditions
$\Gamma_\alpha=D_\alpha/F\gg1$ hold in general and a scaling analysis
similar to that for mono-component adsorbates can be worked out. The
simplest situation arises when all dimers are stable. This corresponds
to the case $i=1$, where
\begin{equation}\label{eq:Gammaeff}
N \simeq \left(\frac{3\sigma_1 \Theta}{\bar\sigma^2 }\right)^{1/3}
 \Gamma_{\rm eff}^{-1/3}\,.
\end{equation}
The exponents agree with those in Eq.~(\ref{eq:N-scaling}) for
one-component systems, but the scaling variable $\Gamma_{\rm
  eff}=D_{\rm eff}/F$ with $D_{\rm eff}=D_A D_B /(x_A D_B + x_B D_A)$
now describes the influence of both the different diffusion
coefficients and mole fractions of the two species. Clearly, the
species with lower diffusion coefficient governs the dependence of $N$
on temperature.

If all dimers are unstable ($i=2$), we first have to seek for
generalized Walton relations~\cite{Einax/etal:2007a} to express dimer
densities in terms of adatom densities. As in
Sec.~\ref{subsec:one-component}, these are obtained by nearly
balancing the formation and dissociation of dimers,
\begin{align}
  \sigma_1 D_\alpha n_\alpha^2&\simeq
  K_{\alpha\alpha}n_{\alpha\alpha}\,,
  \hspace{2em}\alpha=A,B\,,\label{eq:walton1a}\\[1ex]
  \sigma_1(D_{\rm A}+D_{\rm B})n_{\rm A}n_{\rm B}&\simeq
  K_{AB}n_{AB}\,.
\label{eq:walton1b}
\end{align}
Writing $K_{\alpha\beta}=\mu_{\alpha\beta}
D_{\alpha\beta}\exp(-E_{\alpha\beta}/k_{\rm B}T)$ in terms of the
diffusion coefficients $D_{\alpha\alpha}=D_\alpha$,
$D_{AB}=(D_A+D_B)/2$, and the dimer binding energies
$E_{\alpha\beta}\ge0$ (with $\mu_{\alpha\beta}$ being constants), one
obtains
\begin{align}
  N&\simeq
  \left(\frac{4\sigma_1\sigma_2\Theta}{\bar\sigma^3}\right)^{1/4}
  \left[\sum_{\alpha,\beta}
    \mu_{\alpha\beta}^{-1}\,e^{\frac{\scriptstyle
        E_{\alpha\beta}}{\scriptstyle k_{\rm B}T}} \,\frac{F_\alpha
      F_\beta}{D_\alpha D_\beta} \right]^{1/4}
\label{eq:N2}
\end{align}
Equations~(\ref{eq:Gammaeff}) and (\ref{eq:N2}) generalize the scaling
relation (\ref{eq:N-scaling}) for $i=1,2$ to the binary case. The
predictions are in excellent agreement with KMC simulations
\cite{Einax/etal:2007a}, but still await experimental confirmation.
Because of the dependence of $N$ on $E_{AB}$ in Eq.~(\ref{eq:N2}),
binding energies between \emph{unlike} atoms on a substrate should
become accessible by experiment. To achieve this, $U_{\alpha}$ and
$E_{\alpha \alpha}$ can first be determined under deposition of only
one species and then $N$ is measured under co-deposition. The
feasibility of a corresponding procedure has been demonstrated by
recovering all growth parameters from a ``computer experiment'' on
island densities, in a way equivalent to analyzing a true experiment
[for details, including consistency checks, see
\textcite{Einax/etal:2009}].

\begin{figure}[t!]
\includegraphics[width=0.45\textwidth,clip=,]{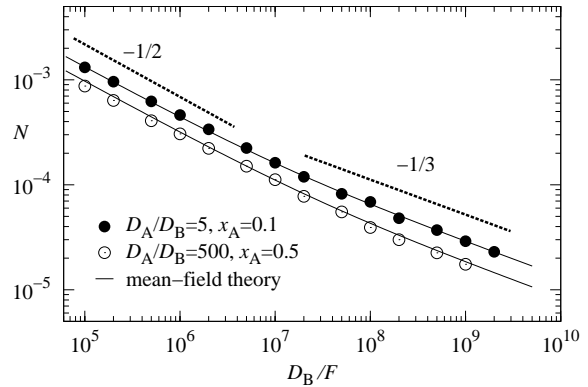}
\caption{Simulated number densities $N$ of stable islands at coverage
  $\Theta=0.1$ as a function of $D_B/F$, when $AA$ dimers are
  stable, while $AB$ and $BB$ dimers are unstable. The solid lines
  correspond to the full solution (\ref{eq:Nfull}) and the dashed
  lines indicate the change of scaling behavior associated with the
  dimer and trimer nucleation routes.
  From~\textcite{Einax/etal:2007a}.}
\label{fig:fig4}
\end{figure}

More complex situations arise, when dimers with different compositions
have different stability. For example, only $BB$ dimers could be
unstable ($K_{AA}=K_{AB}=0$, $K_{BB}>0$), or only $AA$ dimers stable
($K_{AB}$, $K_{BB}>0$, $K_{AA}=0$). For these and similar situations a
closed rate equation for $N$ can be derived by factorizing dimer
densities in terms of the adatom densities with the help of
Eqs.~(\ref{eq:walton1a}, \ref{eq:walton1b}), and by expressing the
adatom densities through $N$ via the quasi-stationary condition
analogous to Eq.~(\ref{eq:n-scaling}). In all cases this leads to
\begin{equation}\label{eq:dNdt1}
 \frac{dN}{d\Theta}=\frac{b}{N^2}+\frac{c}{N^3}\,.
\end{equation}
The linear combination on the right hand side reflects nucleation via
the dimer ($dN/d\Theta\sim N^{-2}$) and trimer route ($dN/d\Theta\sim
N^{-3}$).

The full solution of Eq.~(\ref{eq:dNdt1}) is given by
\begin{equation}
\label{eq:Nfull}
N=(c/b)\,\psi(b^4 \Theta/ c^3)\,,
\end{equation}
where $\psi(.)$ is determined by a transcendental equation
\cite{Einax/etal:2007a}. The coefficients $b$ and $c$ in
Eq.~(\ref{eq:dNdt1}) depend on $\Gamma_{\alpha}$ and $x_{\alpha}$, and
their detailed form is given in \textcite{Dieterich/etal:2008}. If,
for example, stable dimers are of $AA$ type only, then $b \propto
F_A/D_A$, and $c$ is a sum of two terms proportional to $(F_A F_B/D_A
D_B)$ and to $(F_B/D_B)^2$, corresponding to the two paths of trimer
formation via AB and BB dimers. Each possible nucleation path in the
above examples contributes additively to $dN/d\Theta$. The dimer
route is dominating for fast $B$ diffusion and $x_A$ large, and the
trimer route for slow $B$ diffusion and $x_B$ large. The smooth
cross-over is validated in Fig.~\ref{fig:fig4}, where KMC data for $N$
versus $D_B/F$ are plotted for a situation with stable $AA$ dimers
($K_{AA}=0$) and zero binding energies of $AB$ and $BB$ dimers.

\subsection{Beyond $\Gamma$ scaling: Self-consistent capture numbers}
\label{subsec:self-consistent}
The efficiency of islands to capture monomers is affected by the
shielding by other islands in their neighborhood. Such shielding
effects can be treated approximately by capture numbers that depend on
the whole set of monomer and island densities~\cite{Bales/Chrzan:1994,
  Bales/Zangwill:1997}. The idea is to consider the diffusion
equation for the local adatom profile $\tilde n(r,t)$ around an
$s$-cluster with radius $R_s=(s/\pi)^{1/2}$ in the presence of the
flux $F$ and an effective medium with absorption rate $D/\xi^2$,
\begin{align}
\label{eq:diffusion_equation_adatom}
\left(\frac{\partial}{\partial t}- D\Delta\right)\tilde n(r,t)&=
F-\frac{D}{\xi^2} \tilde n(r,t)\,,
\end{align}
where $D/\xi^2$ is identified with the loss terms in
Eq.~(\ref{eq:monomer_density}),
\begin{align}
\label{eq:xi}
\xi^{-2}= 2\sigma_1 n+\sum_{s\ge 2}\sigma_s n_s\,.
\end{align}
The sum over $s$ can be replaced by $\bar\sigma N$ as before, leading
to a reduced self-consistent description already considered by
\textcite{Venables:1973}. In a quasi-stationary state,
$\partial_t\tilde n\simeq0$, $D\xi^{-2}n\simeq F$, and
Eq.~(\ref{eq:diffusion_equation_adatom}) becomes
\begin{align}
\label{eq:helmholtz_equation}
\Delta\tilde n(r) - \frac{1}{\xi^2}\left[\tilde n(r)- n\right] &=0 \, .
\end{align}
For large $r$, $\tilde n(r)$ should approach its mean value $n$
considered in the rate equations and at the island edge $r=R_s$, it
must be zero in the absence of detachment processes (case $i=1$; for a
generalization to higher $i$, see \textcite{Bales/Zangwill:1997}).
With these boundary conditions, the solution of
Eq.~(\ref{eq:helmholtz_equation}) becomes $\tilde
n(r)=n[1-K_0(r/\xi)/K_0(R_s/\xi)]$, where $K_\nu(.)$ denotes the
modified Bessel function of order $\nu$. Equating the inward current
of adatoms $2\pi R_s D \left( \partial\tilde n/\partial r
\right)_{r=R_s}$ at the island edge with the aggregation rate
$\sigma_s D n$, one obtains
\begin{align}
\label{eq:flux_matching-boundary_condition}
\sigma_s = \frac{2 \pi R_s}{n} \left(\frac{\partial\tilde n}{
    \partial r} \right)_{r=R_s}=
2 \pi \frac{R_s}{\xi} \frac{K_1 \left( R_s/\xi \right)}{K_0
\left( R_s/\xi \right) } \, .
\end{align}
The merit of this approach is that a good quantitative description of
$n$ and $N$ as a function of $\Theta$ is provided despite of the fact
that the $\sigma_s$ themselves are not well predicted, as discussed
later [cf.\ Fig.~\ref{fig:isd}(a)].

For a generalization of the self-consistent theory to multi-component
systems, many-particle densities need to be introduced to obtain
expressions for capture numbers that are symmetric under exchange of
atomic species \cite{Einax/etal:2012}. The method is best illustrated
by considering the reaction rate $(D_A+D_B)\sigma_1^{AB}n_An_B$ of $A$
and $B$ monomers to form $AB$ dimers in the case of binary alloys with
$i=1$. The effective medium in this case is characterized by two mean
free paths $\xi_{\alpha}$, $\alpha=A,B$, which result from the loss
terms $(D_{\alpha}/\xi_{\alpha}^2)\,n_{\alpha}$ in the rate equations
of the monomer densities. Considering the pair distribution function
of $A$ and $B$ monomers, this function satisfies a diffusion equation
with symmetric diffusion coefficient $(D_A+D_B)$ and symmetric
absorption rate $\sum_{\alpha} D_{\alpha} \xi_{\alpha}^{-2}$. Taking
into account the boundary conditions, the final result for
$\sigma_1^{AB}$ has the same structure as in the one-component
theory~\cite{Bales/Chrzan:1994},
\begin{align}
\label{eq:flux_matching-boundary_condition2}
\sigma_1^{AB} &= 2 \pi \frac{R_1}{\xi_{\rm{eff}}}
\frac{K_1 \left( R_1/\xi_{\rm{eff}} \right)}{K_0
\left( R_1/\xi_{\rm{eff}} \right) }
\end{align}
with an effective mean free path $\xi_{\rm{eff}}$.  In contrast to the
one-component theory this mean free path
depends on the diffusion coefficients,
\begin{align}
\label{eq:xieff}
\xi_{\rm{eff}}^{-2} = (D_A+D_B)^{-1}[D_A \xi_A^{-2} +D_B \xi_B^{-2}]\,.
\end{align}
One can show that $\sigma_1^{\alpha\alpha}$ is also given by
Eq.~(\ref{eq:flux_matching-boundary_condition2}) with $\xi_{\rm{eff}}$
replaced by $\xi_{\alpha}$. To get $\sigma_s^\alpha$ for $s>1$ one has
in addition to replace $R_1$ by $R_s$.

\begin{figure}[t!]
\includegraphics[width=0.45\textwidth,clip=,]{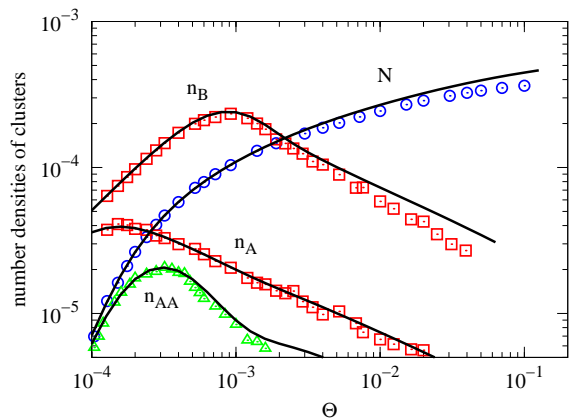}
\caption{(Color online) Number densities of $A$ and $B$ adatoms, $AA$
  dimers and stable islands as a function of the coverage $\Theta$ for
  a case of mixed dimer stabilities, where $AA$ dimers are stable,
  while $AB$ and $BB$ dimers are unstable with zero binding
  energies. Results from the self-consistent approach (solid lines)
  are compared with KMC simulations (symbols).  Parameters are
  $x_A=x_B=1/2$, $D_A/F=10^8$, $D_B/F=10^7$.}
\label{fig:fig5}
\end{figure}

As in the mono-component case, an extension is possible also to
include decay processes of clusters. The capture and decay rates in
this extension satisfy the generalized (two-component) Walton
relations. Again a good quantitative description of island densities
is obtained. As an example we show in Fig.~\ref{fig:fig5} calculated
number densities $N$, $n_{\alpha}$ and $n_{\alpha \alpha}$ for mixed
dimer stabilities in comparison with KMC data.

For critically testing the theory outlined here, it would be important
to provide experimental data for island densities in multi-component
systems. A particular challenge is to measure simultaneously the
densities of the individual monomers or more generally, the densities
of the subcritical clusters.  This is already highly desirable for
mono-component systems, because it allows one to test basic
assumptions of the rate equation approach. For example, to our
knowledge, neither the Walton relation (\ref{eq:Walton_relation}) nor
their generalized forms (\ref{eq:walton1a}) and (\ref{eq:walton1b})
have been tested in experiments.

To sum up, using rate equations and KMC simulations, it was shown that
even for binary systems comparatively simple (multi-parameter) scaling
relations can be established that describe the densities of stable
islands as a function of adatom diffusion coefficients and partial
deposition fluxes.
Equation~(\ref{eq:flux_matching-boundary_condition}) entails a
generalization to binary systems of the well-known ``nucleation
route'' for obtaining microscopic parameters from island density
measurements. In practice, this equation can be utilized to determine
binding energies between unlike atoms in the presence of the surface.
It would be interesting to compare these results with corresponding
ones from electronic structure calculations. A particularly
interesting feature is the occurrence of novel cross-over phenomena in
the island densities, which emerge when critical nuclei of different
composition compete with each other in determining the dominant
nucleation pathway. A detailed quantitative description of island
densities is possible, when considering many-particle densities in
self-consistent theories for capture numbers. Also decay processes of
unstable clusters can be treated successfully within this framework.

\section{What is the distribution of island sizes?}
\label{sec:size_distribution}

\subsection{Predictions from rate equations}
\label{subsec:RE_predictions}
More detailed information of the submonolayer growth kinetics is
contained in the island size distribution (ISD) $n_s(\Theta,\Gamma)$.
Imagine that the capture numbers $\sigma_s$ in Eqs.~(\ref{eq:rate1})
were known. Can we then expect that the RE predict the ISD in the
pre-coalescence regime? This would mean that many-particle correlation
effects can be incorporated in effective capture numbers, and that
spatial fluctuations in shapes and capture zones of islands as well as
coalescence events, despite rare in the early-stage growth, are
negligible.

\begin{figure}[t!]
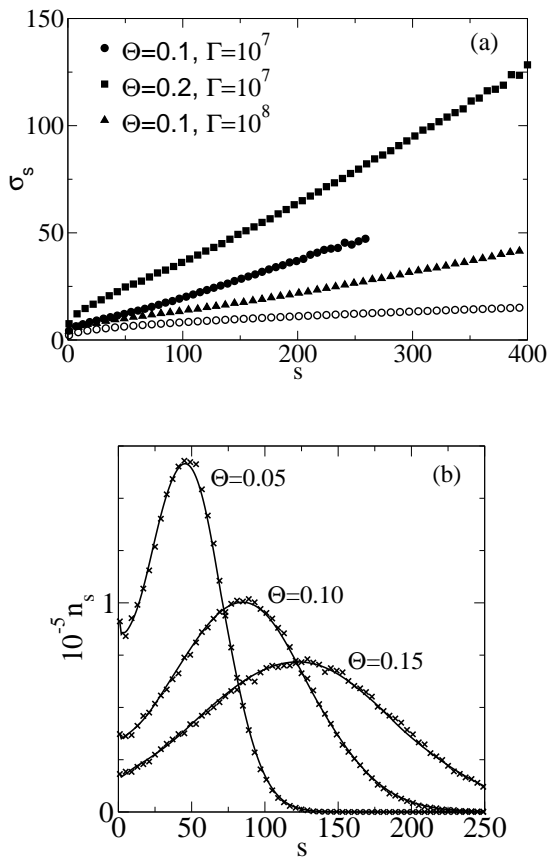

\includegraphics[width=0.4\textwidth,clip=,]{fig6a}\\[6ex]
\includegraphics[width=0.34\textwidth,clip=,]{fig6b}\\[0ex]
\caption{(a) Capture numbers and (b) island size distributions from
  KMC simulations of kinetic growth with hit-and-stick aggregation on
  a (100) surface. In (a) the open circles refer to the prediction of
  the self-consistent theory for $\Theta=0.1$ and $\Gamma=10^7$, which
  for large $s$ deviate strongly from the true
  $\sigma_s(\Theta,\Gamma)$ (full circles). In (b) the solid lines are
  the solution of the RE~(\ref{eq:rate1}) with the simulated
  $\sigma_s(\Theta,\Gamma)$ as input. From~\textcite{Koerner/etal:2012}.}
\label{fig:isd}
\end{figure}

While Eq.~(\ref{eq:flux_matching-boundary_condition}) already implies
that capture numbers have to regarded as effective ones,
$\sigma_s=\sigma_s( \Theta, \Gamma)$, their full dependence on both
island size $s$ and external parameters has been determined recently
in an extensive KMC study of models yielding different island
morphologies \cite{Koerner/etal:2010,Koerner/etal:2012}. In
Fig.~\ref{fig:isd}(a) we show representative results of this study for
kinetic growth with hit-and-stick aggregation on a (100) surface.
Integrating the refined RE [see discussion after
Eqs.~(\ref{eq:rate1})] under consideration of the full
$\Theta$-dependence of $\sigma_s(\Theta,\Gamma)$ indeed gives a very
good description of the ISD, as demonstrated in Fig.~\ref{fig:isd}(b).
To achieve this good agreement it is necessary to take into account
details in the functional form of $\sigma_s(\Theta,\Gamma)$. For
example, when neglecting the $\Theta$-dependence by setting
$\sigma_s(\Theta,\Gamma)=\sigma_s(\Theta_0,\Gamma)$ for a fixed
reference coverage $\Theta_0$, a good description is not obtained.  It
is moreover important to point out that for compact island
morphologies the good agreement is of limited practical use, because
coalescence events, not considered in the RE, become relevant already
at rather small coverages of about 5\%.

Since the refined RE with those simulated $\Theta$-dependent
$\sigma_s$ were shown to be suitable for predicting the ISD the
challenge is to find an accurate description of the functional form of
the $\sigma_s(\Theta,\Gamma)$. Unfortunately the
$\sigma_s(\Theta,\Gamma)$ show a high sensitivity with respect to the
island morphologies and probably also to details of the growth
kinetics.  A general feature is that for island sizes $s$ larger than
the mean island size $\bar s$, $\sigma_s(\Theta,\Gamma)$ increases
linearly with $s$, $\sigma_s(\Theta,\Gamma)\sim
c_1(\Theta,\Gamma)+c_2(\Theta,\Gamma)s$. This can be reasoned by
considering the areas for adatom capture surrounding the islands
\cite{Evans/etal:2006}. However, even for this generic feature the
functions $c_j(\Theta,\Gamma)$ depend on details of the attachment
kinetics (hit-and-stick or with limited edge diffusion, etc.). For
improving theories for the ISD, it has been argued that correlation
effects between island sizes and capture areas need to be taken into
account. This can been achieved by considering the joint probability
of island size and capture area \cite{Amar/etal:2001,
  Popescu/etal:2001, Mulheran/Robbie:2000, Evans/Bartelt:2002,
  Mulheran:2004}.

\subsection{Scaling approaches}
\label{subsec:isd-scaling}
While a theoretical description of the details in
$\sigma_s(\Theta,\Gamma)$ seems to be out of reach, a simpler approach
to the ISD becomes possible when assuming that the dependence of
$n_s(\Theta,\Gamma)$ on $\Theta$ and $\Gamma$ is mediated by the mean
island size $\bar s(\Theta,\Gamma)$. In this case the ISD should obey
the following scaling form, as first suggested by
\textcite{Vicsek/Family:1984}
\begin{align}
\label{eq:scaling_form}
n_s (\Theta,\Gamma)= \frac{ \Theta }{ {\overline s}^2 (\Theta, \Gamma) }
f\left(\frac{s}{\bar s(\Theta,\Gamma)}\right) \;\;.
\end{align}
Here the scaling function $f(x)$ must fulfill the normalization and first moment conditions
$\int_0^{\infty} f(x) dx = \int_0^\infty x f(x) dx =1$, because
$\sum_s n_s\simeq N\simeq \Theta/\bar s$ and $\sum_s sn_s=\Theta$.

An explicit expression for $f(x)$ with a shape independent of $\Theta$
was suggested by \textcite{Amar/Family:1995},
\begin{align}
\label{eq:empirical_form}
f(x) = C_i x^i \exp \left(- i a_{i} x^{1/a_i} \right) \, ,
\end{align}
where the parameters $C_i$ and $a_i$ follow from the
above conditions on $f$.
Equation~(\ref{eq:empirical_form}) was believed to be independent even
of the morphology \cite{Amar/Family:1995},
but this has later been questioned \cite{Bartelt/Evans:1996, Evans/etal:2006}.
The dependence of Eq.~(\ref{eq:empirical_form}) on $i$ allows one to determine
the size of the critical nucleus from measurements of the ISD
\cite{Loske/etal:2010,Potocar/etal:2011, Ruiz/etal:2003, Pomeroy/Brock:2006}.

It is interesting to note that a semi-empirical form, which has a
structure similar to Eq.~(\ref{eq:empirical_form}) has been suggested
by \textcite{Pimpinelli/Einstein:2007} for the distribution of capture
zone areas $A$ as identified by Voronoi tessellation,
\begin{align}
\label{eq:Capture_zone_distribution}
P_\beta &= c_\beta a^\beta \exp(-d_\beta a^2) \, ,
\end{align}
where $a =A/\bar A$ is the capture zone rescaled with respect to the
mean $\bar A$ and $\beta = i + 2$
\cite{Li/etal:2010,Pimpinelli/Einstein:2010}.\footnote{In the original
  work \cite{Pimpinelli/Einstein:2007} $\beta = i+1$ was suggested
  [see also \cite{Oliveira/Reis:2011,Oliveira/Reis:2012}].} The
parameters $c_\beta$ and $d_\beta$ are again determined by
normalization and a conditions on the first moment.

This new theoretical result can alternatively be used to determine
$i$. In fact, the capture-zone scaling has recently been applied to
determine critical island sizes in organic thin film growth
\cite{Lorbek/etal:2011,Potocar/etal:2011,Tumbek/etal:2012}.  For
para-hexaphenyl (6P) molecules deposited on a sputter-modified
muscovite mica (001) substrate, \textcite{Potocar/etal:2011} reported
that the combined analysis of the ISD according to
Eq.~(\ref{eq:empirical_form}) and the capture-zone distribution
(\ref{eq:Capture_zone_distribution}) gave convincing evidence for
$i=3$. More experiments in this direction should be performed to test
the general validity of Eq.~(\ref{eq:Capture_zone_distribution}). In
particular, the analysis should be extended to identify correlations
between island size and capture zone area. As mentioned above (see
end of Sec.~\ref{subsec:RE_predictions}), corresponding information is
highly relevant to check theoretical considerations on the
size-dependence of capture numbers.

\subsection{Limiting behavior for $D/F \rightarrow \infty$}
\label{subsec:Limiting_behavior}
Because the refined RE with the appropriate $\sigma_s(\Theta,\Gamma)$
predict the ISD, it should be possible to derive an evolution equation
for the scaled island density $f(x,\Theta,\Gamma)\equiv {\bar
  s}^2n_{x\bar s}/\Theta$. Due to the normalization this function will
approach a limiting curve $f_\infty(x,\Theta)$ for $\Gamma\to\infty$,
and an interesting question is whether this curve is independent of
$\Theta$. To discuss this, we concentrate on the case $i=1$.  For
large $\Gamma$, $\bar s\sim N^{-1}\sim\Gamma^{1/3}$, and $x=s/\bar s$
becomes a continuous variable. This allows one to consider a continuum
version of the refined RE \cite{Bartelt/Evans:1996,
  Evans/Bartelt:2001, Koerner/etal:2012}, which gives a partial
differential equation for $f(x;\Theta,\Gamma)$ in the variables $x$
and $\Theta$. This equation includes the scaled capture numbers
$C(x,\Theta,\Gamma)=\sigma_{x\bar s}/\bar\sigma$ and scaled areas for
direct capture $K(x,\Theta,\Gamma)=\kappa_{x\bar s}/\bar\kappa$.
Taking the $\Gamma\to\infty$ limit then yields a determining equation
for $f_\infty(x,\Theta)$. If $f_\infty$ is independent of $\Theta$,
this equation becomes an ordinary differential equation that can be
solved exactly \cite{Bartelt/Evans:1996} and yields a power-law decay
for large $x$.  However, arguments presented by
\textcite{Oliveira/Reis:2012} support a stretched exponential decay of
the scaling function.

To study this question of a $\Theta$ dependence of
$f_\infty(x,\Theta)$ for different island morphologies in the case
$i=1$, one can use the simulated capture numbers $\sigma_s (\Theta)$,
discussed in Sec.~\ref{subsec:RE_predictions}
\cite{Koerner/etal:2012}. In addition, the connection of
$f_\infty(x,\Theta)$ to the scaled capture numbers and areas in the
limit $\Gamma\to\infty$ was studied.  For the point island model
\cite{Bartelt/Evans:1992}, where islands have no extension and
$\Theta=Ft$ plays the role of the deposition time, $f_\infty$ indeed
turned out to independent of $\Theta$, while for extended islands the
results indicate that $f_\infty$ exhibits a (weak) $\Theta$
dependence.

To conclude taking into account the full dependence of capture numbers
on island size, coverage and $D/F$-ratio, the rate equation approach
predicts well the ISD in the pre-coalescence regime. Moreover,
analysis of ISDs and capture cone area distributions allow for a
determination of $i$, independent of the methods presented in
Sec.~\ref{sec:islands}. Further theoretical work is required in order
to establish the behavior of the limiting curve $f_\infty (x)$.

\section{When do islands survive upon continued deposition?}
\label{sec:second_layer}
With ongoing deposition, when multilayer growth sets in, two
contrasting growth modes can arise. Islands in the first layer either
grow in the lateral direction and coalesce before a new layer
nucleates, or stable nuclei form on top of them before coalescences.
Which of these two cases is realized, depends on the characteristic
island radius $R_c$ at the onset of second layer nucleation: If $R_c$
is larger than the mean distance $N^{-1/2}$ between islands,
coalescence sets in before second layer nucleation and a smooth
layer-by-layer growth behavior is obtained. On the other hand, if
$R_c$ is smaller than $N^{-1/2}$, islands grow in the normal direction
before a smooth layer has developed and the film topography becomes
rough.

It is important to note that the occurrence of smooth or rough films
is answered differently from a thermodynamic viewpoint, where it is
dictated by surface tensions and associated wetting properties. In
this context a rough Volmer-Weber, a smooth Van-der-Merwe and an
intermediate Stranski-Krastanov growth mode are distinguished
~\cite{Pimpinelli/Villain:1998}.
During growth, however, films are usually not in thermal
equilibrium and the film topography is determined by kinetics
rather than thermodynamics.
Under non-equilibrium growth conditions, $R_c$ is the
decisive quantity that determines the film topography
\cite{Tersoff/etal:1994, Rottler/Maass:1999}.

An important quantity controlling the size of $R_c$ is the
Ehrlich-Schwoebel or additional step edge barrier $\Delta
E_{\scriptscriptstyle\rm ES}$ that an atom has to surmount when
passing an island edge. If $\Delta E_{\scriptscriptstyle\rm ES}$ is
large, the associated Boltzmann factor
\begin{equation}
W=\exp(-\Delta E_{\scriptscriptstyle\rm ES}/k_{\rm B}T)
\label{eq:ws}
\end{equation}
becomes small and atoms can remain longer on top of islands, implying
that second layer nucleation becomes more likely.\footnote{The step
  edge barrier can be an effective one, i.e.\ $\Delta E_{\scriptscriptstyle\rm ES}=-k_{\rm
    B}T\ln\langle\exp(-\Delta E_\alpha/k_{\rm B}T)\rangle$, where
  $\langle\ldots\rangle$ denotes an average over microscopic step edge
  barriers $\Delta E_\alpha$ for different local atomic configuration
  $\alpha$ at the island edge. Moreover, a factor $\nu_s/\nu_{\rm
    isl}$ needs to be included in Eq.~(\ref{eq:ws}) if the attempt
  frequency $\nu_s$ for hops down the step edge is different from the
  attempt frequency $\nu_t$ for jumps between sites on top of
  the island.}

\subsection{Second-layer nucleation rate}
\label{subsec:second_layers}
In order to develop a theory, which predicts how $R_c$ depends on $W$
and $\Gamma$, a precise definition of $R_c$ is needed. More
generally, consider at time $t$ the fraction $f(t)$ of islands covered
by stable clusters. This fraction is zero for small $t$ and increases
to one in some time window, after which all islands are covered. At a
time $t_c$ half of the islands are covered, $f(t_c)=0.5$, and the
critical radius $R_c$ can be conveniently defined by $R_c=R(t_c)$,
where $R(t)$ denotes the mean island radius at time $t$.
Let us define by $\Omega(R)$ the second-layer nucleation rate. The
fraction of covered islands then is \cite{Tersoff/etal:1994}
\begin{equation}
f(t)=1-\exp(-\int_0^t {\rm d}t'\, \Omega(R(t'))\,,
\label{eq:f(t)}
\end{equation}
and the problem reduces to determine $\Omega(R)$ in dependence of $W$
and $\Gamma$.

The functional form of $\Omega(R)$ can be deduced from scaling
arguments \cite{Rottler/Maass:1999, Heinrichs/etal:2000,
  Krug/etal:2000, Krug:2000}.  To this end we consider a state, where
in total $n$ adatoms are simultaneously on top of an island of size
$R$. A number of $(i+1)$ adatoms can encounter at a point of the
island with a probability $\sim (a^2/\pi
R^2)^{(i+1)}\prod_{k=0}^i(n-k)$, where the product takes into account
that there are $n(n-1)/2$ possibilities to form an (intermediate)
pair, and, for $i>1$, $(n-2)\times(n-3)\ldots\times(n-i)$ further
possibilities for the remaining $(i-1)$ adatoms to attach to the pair.
Multiplying this probability with the adatom diffusion rate $D/a^2$
and integrating over the island area (factor $\pi R^2$) yields the
encounter rate of $(i+1)$ atoms in the presence of $n\geq (i+1)$ atoms
on an island of size $R$:\footnote{To keep the notation simple, we
  have not introduced a new coefficient $D'$ for adatom diffusion in
  the second layer, which, of course, can be different from $D$ in the
  first layer.}
\begin{align}
  \omega_n(R)&=\kappa_e \frac{D}{a^2}
\left[\prod_{k=0}^i(n-k)\right]\left(\frac{a^2}{\pi R^2}\right)^i\,.
\label{eq:encounter_rate}
\end{align}
Logarithmic corrections need to be included in this expression when
taking into account that the encounter problem involves the number of
distinct sites visited by a diffusing adatom
\cite{Politi/Castellano:2003}.

The decay time $\tau_n(R)$ of a state with $n$ non-interacting adatoms
on top of an islands is the $n$th fraction of the lifetime $\tau_1(R)$
of one adatom. The latter is given by the typical time $\sim R^2/D$ to
reach the island edge plus the time $\sim Ra/D$ to return to the edge,
which on average is taking place $W^{-1}$ times before the step
edge barrier is eventually passed (if no second layer nucleation
occurs). Accordingly,
\begin{align}
  \tau_n(R)&=\frac{1}{n}\tau_1(R)=
\frac{1}{n}\left(\kappa_1\frac{R^2}{D}+\kappa_2\frac{Ra}{W D} \right)\,.
\label{eq:lifetime}
\end{align}
Equations~(\ref{eq:encounter_rate}) and (\ref{eq:lifetime}) have been
validated by more detailed analytical calculations and tested against
KMC simulations in \textcite{Heinrichs/etal:2000}, where in particular
the constants $\kappa_1$ and $\kappa_2$ were determined and $\kappa_e$
was shown to depend on $i$ due to memory effects.

Knowledge of the $\tau_n(R)$ allows us to calculate the probabilities
$p_n=p_n(R(t))$ of finding $n$ adatoms on top of the island at a time
$t$ before the onset of second layer nucleation. These are given by
the Poisson distribution
\begin{align}
p_n(R)=\frac{\bar n(R)^n}{n!}\exp[-\bar n(R)]\,,
\label{eq:pn}
\end{align}
where an analytical expression for the mean number $\bar n=\bar
n(R(t))$ of adatoms was derived in \textcite{Heinrichs/etal:2000}. It
holds $\bar n(R)=\pi FR^2\tau_1(R)$, except for a regime of small
$R/a\ll W^{-1}\Gamma^{-2/(i+2)}$.

With the encounter rates $\omega_n(R)$, the lifetimes $\tau_n(R)$, and
the probabilities $p_n(R)$, the second-layer nucleation rate can be
evaluated. Different from what one may intuitively expect, it is
possible that second layer nucleation sets in at times when $\bar
n(R(t))\ll (i+1)$, that means when on average less than $(i+1)$
adatoms are on top of the island \cite{Rottler/Maass:1999}. In this
case the nucleation is caused by fluctuations, where by chance $(i+1)$
adatoms are on top of an island and encounter each other to form a
stable nucleus.

To calculate the corresponding fluctuation-dominated nucleation rate
$\Omega_{\rm fl}(R)$, consider the probability of a second layer
nucleation event in a finite time interval $\Delta t$.\footnote{The
  interval $\Delta t$ should be large compared to $\omega_{i+1}^{-1}$
  and small compared to times scales of changes of $R(t)$.} This
equals the probability $p_i(R(t))$ of finding $i$ atoms on top of the
island (states with $n>i$ can be neglected in the
fluctuation-dominated situation), times the probability $\pi
FR^2\Delta t$ to deposit an additional atom on the island, times the
probability $(1-\exp[-\omega_{i+1}(R)\tau_{i+1}(R)])$ that $(i+1)$
adatoms encounter each other during the lifetime
$\tau_{i+1}(R)$. Dividing the probability of a nucleation event in
$\Delta t$ by $\Delta t$ yields
\begin{align}
\Omega_{\rm fl}(R)&=
\pi FR^2 \frac{\bar n(R)^i}{i!}e^{-\bar n(R)}
\Bigl(1-e^{-\omega_{i+1}(R)\tau_{i+1}(R)}\Bigr)\,.
\label{eq:om-fl}
\end{align}

For $i=1$, Eqs.~(\ref{eq:encounter_rate}) and (\ref{eq:lifetime}) give
$\omega_2(R)\tau_2(R)\sim (1/RW)$ for $RW\ll1$ and
$\omega_2(R)\tau_2(R)\sim const.$ for $RW\gg1$, which means that the
encounter probability $(1-\exp[-\omega_2(R)\tau_2(R)])$ in
Eq.~(\ref{eq:om-fl}) is always of order one. One thus obtains
\begin{align}
\Omega_{\rm fl}(R)&\sim
FR^2\bar n(R)\sim F^2R^4\tau_1(R)\nonumber\\
&\sim\left\{\begin{array}{l@{\hspace*{0.2cm}}l}
FW^{-1}\Gamma^{-1}R^5\,, &
        \Gamma^{-11/12} \ll W R\ll 1\\[0.2cm]
F\Gamma^{-1}R^6\,, & 1\ll W R
\end{array}\right.
\label{eq:om-fl-i1}
\end{align}
For $W R\ll \Gamma^{-11/12}$ it holds $\Omega_{\rm fl}(R)\sim
F\Gamma^{-1/3}R^6$.

If $\bar n(R(t))\gtrsim (i+1)$ at the onset of second layer
nucleation, the rate follows from the weighted sum of the encounter
rates $\omega_n(R)$ over all states with $n\ge(i+1)$,
\begin{align}
\Omega_{\rm mf}(R)&=\sum_{n=i+1}^\infty
p_n(R)\omega_n(R)=\kappa_e\frac{D}{a^2}\bar n(R)^{i+1}
\left(\frac{a^2}{\pi R^2}\right)^i\,.
\label{eq:om-mf}
\end{align}
This agrees with the result of mean-field theory
\cite{Tersoff/etal:1994}, which predicts that $\Omega(R)$ is given by
an integration of the local nucleation rate $Dn_1^{i+1}\sim D(\bar
n(R)/\pi R^2)^{i+1}$ over the island area (factor $\pi R^2$) .

By a self-consistent analysis it can decided which of the two
different situations leading to Eqs.~(\ref{eq:om-fl}) and
(\ref{eq:om-mf}) actually occurs. Assume that the
fluctuation-dominated situation is relevant. Then, calculating $f(t)$
from Eq.~(\ref{eq:f(t)}) with $\Omega(R)=\Omega_{\scriptscriptstyle\rm
  fl}(R)$ from Eq.~(\ref{eq:om-fl}), and determining $R_c$, one can
check if $\bar n(R_c)\ll (i+1)$. If this is true, the assumption of a
fluctuation-dominated nucleation was correct, while otherwise the
mean-field rate from Eq.~(\ref{eq:om-mf}) must be used. When $R(t)$
evolves according to the natural growth law
$(FN^{-1}t)^{1/2} \sim F^{1/2} \Gamma^{i/2(i+2)} t^{1/2}$,
it turns out that for $i\le2$ the fluctuation-dominated rate $\Omega_{\rm fl}(R)$
is relevant, while for $i>2$ it is the mean-field rate $\Omega_{\rm
  mf}(R)$ \cite{Heinrichs/etal:2000}.

With respect to the dependence of $R_c$ on $W$ and $\Gamma$ the theory
based on Eq.~(\ref{eq:om-fl}) predicts various scaling regimes in
different intervals of $W$, where $R_c\sim W^\eta\Gamma^\gamma$ with
exponents $\eta$ and $\gamma$ depending on $i$.  For example, for
$i=1$, one finds $R_c\sim W^{1/7}\Gamma^{4/21}$ for $W^{(1)}\ll W\ll
W^{(2)}$, where $W^{(1)}\sim\Gamma^{-3/4}$ and
$W^{(2)}\sim\Gamma^{-1/6}$.  The occurrence of the different regimes
and their scaling properties were confirmed by KMC simulations
\cite{Rottler/Maass:1999,Heinrichs/etal:2000}.

In extensions of this theory, also more complicated cases of second
layer nucleation can be treated, including the presence of metastable
clusters and of interaction effects between adatoms
\cite{Heinrichs/etal:2000, Heinrichs/Maass:2002}. Moreover, the same
type of fluctuation effects as discussed here lead to a failure of
mean-field rate equations for describing chemical reaction kinetics in
confined geometries, as, for example, hydrogen recombination on
interstellar dust grains. Similar scaling arguments can be used to
tackle this related problem \cite{Krug:2003}.

\subsection{Applications}
\label{subsec:Appl_second_layer}
The theory of second layer nucleation allows one to determine phase
diagrams, where in dependence of the two parameters $\Gamma$ and $W$
it is predicted whether films grow into smooth or rough topographies.
While such phase diagrams have been validated by KMC simulations
\cite{Rottler/Maass:1999}, corresponding experimental studies are
still lacking. For a given adsorbate and substrate, different paths in
the $W$-$\Gamma$-diagram could be explored by changing the flux $F$
and the temperature $T$. Care would be needed in analyzing such
studies because changes in $i$ can go along with changes of $T$.

\begin{figure}[t!]
\includegraphics[width=0.4\textwidth,clip=,]{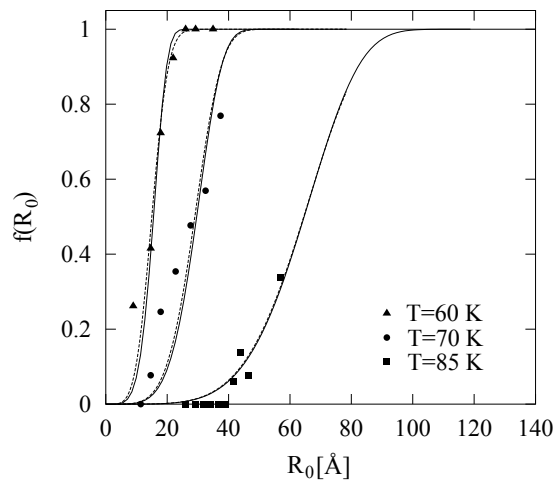}
\caption{Fraction of covered islands after deposition of 10\% of a
  monolayer of Ag on a Pt(111) surface with size-tailored islands of
  mean radius $R_0$ at three different temperatures ($i=1$). The
  symbols refer to the measured data by \textcite{Bromann/etal:1995},
  and the solid and dashed lines (almost identical) refer to fits with the second-layer
  nucleation theory under neglect [Eqs.~(\ref{eq:om-fl}) or
  (\ref{eq:om-fl-i1})] and inclusion of additional ring barriers
  between mutually approaching adatoms, respectively. From
  \textcite{Heinrichs/Maass:2002}.}
\label{fig:fig7}
\end{figure}

A further application of second-layer nucleation theory is the
determination of step edge barriers from measurements of the fraction
of covered islands. This supplements other techniques as field ion
microscopy \cite{Kellogg:1994}, which can suffer from the problem that
step edges are not resolved with a sufficient resolution. With STM
\textcite{Bromann/etal:1993} have analyzed the fraction of covered
islands in homo- and heteroepitaxial Ag growth on Ag(111) and Pt(111).
They first generated size-tailored Ag islands with a narrow size
distribution around a mean radius $R_0$ by deposition of 10\% of a
monolayer and subsequent annealing. Different $R_0$ in the range
10-100~\AA\ were obtained in dependence of the annealing time. After a
subsequent evaporation of again 10\% of a monolayer at different
temperatures, the fraction of covered islands $f(R_0)$ as a function
of the initial radius $R_0$ was determined, as shown in
Fig.~\ref{fig:fig7} for Ag growth on Pt(111). In this system $i=1$
below 90~K \cite{Brune/etal:1994}.  When assuming $\nu_s=\nu_t$, an
analysis of the measured $f(R_0)$ with the fluctuation-dominated rate
$\Omega_{\rm fl}(R)$ from Eq.~(\ref{eq:om-fl-i1}) yields $\Delta
E_{\scriptscriptstyle\rm ES}\simeq43$~meV, while the mean-field rate
$\Omega_{\rm mf}(R)$ would give a too small value of $\Delta
E_{\scriptscriptstyle\rm ES}\simeq30$~meV. When taking into account
interactions in form of additional ring barriers between mutually
approaching adatoms, as predicted by DFT calculations of
\textcite{Fichthorn/Scheffler:2000}, a value $\Delta
E_{\scriptscriptstyle\rm ES}\simeq48$~meV is obtained (for details,
see \textcite{Heinrichs/Maass:2002}).

\begin{figure}[b!]
\includegraphics[width=0.3\textwidth,clip=,]{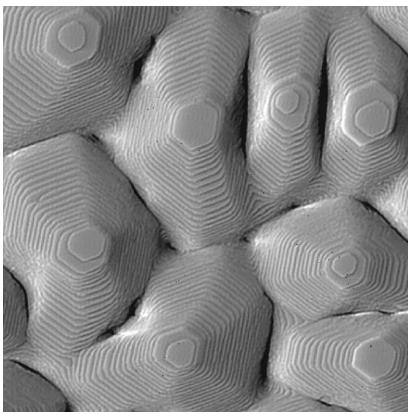}
\caption{STM image of mound formation of Pt after deposition on
  Pt(111). From \textcite{Krug/etal:2000}.}
\label{fig:fig8}
\end{figure}

In the case of rough multilayer growth, an alternative way for
determining $\Delta E_{\scriptscriptstyle\rm ES}$ is to measure the
mean radius $R_{\rm top}$ of the top terrace after mound formation in
the so-called ``wedding cake regime'' \cite{Politi:1997}. During mound
growth, see Fig.~\ref{fig:fig8}, the rate $F$ of creating a new layer
should balance the rate of creating a new nucleus on the top terrace,
which yields $\Omega(R_{\rm top})\simeq F$ as a simple determining
equation for $R_{\rm top}$.  \textcite{Krug/etal:2000} refined this
argument, developed an analytical theory for the size distribution of
top terraces and applied this theory to the mound formation of Pt on
Pt(111) shown in Fig.~\ref{fig:fig8}.

A strength of the theory of second layer nucleation is that it does
not depend on microscopic details of the interlayer transport.  What
counts is that there is an {\it effective} barrier $\Delta
E_{\scriptscriptstyle\rm ES}$ that hinders the escape of particles
from an island. This generality was nicely demonstrated by
\textcite{Hlawacek/etal:2008} who applied the theory to mound
formation of rodlike para-sexiphenyl molecules in organic thin-film
growth. Upon deposition on an ion-bombarded mica surface these
molecules are going to stand upright on the surface with some tilt
angle. By analyzing $R_{\rm top}$ for films with large thicknesses a
value $\Delta E_{\scriptscriptstyle\rm ES}\simeq0.67$~eV was
determined by using the stochastic theory with $i=1$ in this case.
Using a molecular model, the authors could show that a large
contribution to this effective barrier stems from the bending of a
molecule when it slides down a step edge, a mechanism very different
from those responsible for $\Delta E_{\scriptscriptstyle\rm ES}$ in
metal epitaxy. For the first molecular layers a lower value $\Delta
E_{\scriptscriptstyle\rm ES}\simeq0.26$~eV was determined based on an
analysis of the critical radius $R_c$. This lower value could be
traced back to a smaller tilt angle of the molecules in the first
layers and an associated lowering of the bending barrier
\cite{Hlawacek/etal:2008}.

Extension of second-layer nucleation theories to multi-component
systems is an open problem. Differing mobilities in the second-layer
and differing step edge barriers of the components are expected to
give rise to interesting new effects. For example, under codeposition
of two species with high and low $\Delta E_{\rm ES}$, the species with
higher $\Delta E_{\rm ES}$ should enrich on top of islands.
Accordingly island compositions will depend on temperature and fluxes.
In later stages of growth this may lead to compositional profiles that
are tunable by experiment.

The essential insight from this section is that for small critical
island sizes $i \leq 2$ second layer nucleation relies on rare
fluctuations in the number of adatoms on top of the island, which
occur within a time domain where $\overline{n}(R(t)) \ll i+1$.  A
stochastic theory capturing theses fluctuations reveals a nucleation
rate differing from mean-field predictions. Fitting that theory to
metal and organic growth experiments can yield pronounced corrections
to the step edge barriers $\Delta E_{\rm ES}$ when compared with
predictions from mean-field theory. The criterion distinguishing
between 3-d island and smooth surface growth will be modified
accordingly.

\section{What determines outer shape and inner structure of
  islands?}
\label{sec:3-D}
At thermodynamic equilibrium, cluster shapes for a given cluster size
are governed by the principle of minimal interfacial free energy,
which involves the familiar Wulff construction \cite{Wulff:1901}.  For
two-dimensional islands on surfaces, it is the step free energy which
enters. Under non-equilibrium conditions, cluster shapes are
controlled by atomic or molecular moves at or near the cluster
surface. When attaching to a cluster, adatoms can encounter many
different environments such as facets, edges, kinks, corners, etc.\
with associated changes of elementary jump energies. This leads to a
large variety of outer shapes and inner configurational arrangements.
In the following we discuss some key mechanisms for this kinetically
controlled structure formation.

\subsection{Island shapes on (111) surfaces}
\label{sec:Anisotropy_in_corner_diffusion}
Shapes of two-dimensional islands on (111) surfaces were studied for
various systems in metal epitaxy, as for Pt/Pt(111)
\cite{Brune/etal:1996}, Ag/Pt(111) \cite{Hohage/etal:1996}, Al/Al(111)
\cite{Ovesson/etal:1999}, Ag/Ag(111) \cite{Cox/etal:2005}, and
Au/Pt(111) \cite{Ogura/etal:2006}. They are widely known and well
understood examples of island structures controlled by non-equilibrium
kinetics.

At high temperature, edge and corner diffusion are generally fast
enough to create sharp island edges (``line facets''). However,
different from what one may intuitively expect, islands typically do
not exhibit hexagonal but triangular shapes, which implies a breaking
of the hexagonal symmetry of the (111) substrate lattice.

\begin{figure}[t!]
\includegraphics[width=0.45\textwidth,clip=,]{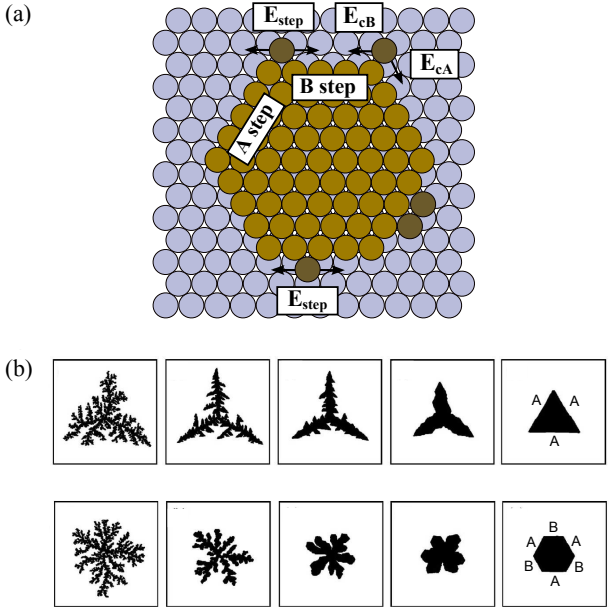}
\caption{(Color online) (a) Sketch of atomic movements and energy
  barriers involved in diffusion processes along and between island
  edges of A and B type on a (111) surface.  (b) Simulated island
  shapes for $T=300$~K, $F=10^{-3}$~ML/s and the barrier for edge
  diffusion $E_{\rm edge}$ decreasing from left to right.  In the
  upper panel, corner diffusion is anisotropic with $E_{\rm
    cB}=0.1\,\mbox{eV}<E_{\rm cA}=0.25\,\mbox{eV}$, while in the lower
  panel, corner diffusion is symmetric with $E_{\rm cB}= E_{\rm
    cA}=0.11$~eV. Adapted from~\textcite{Ogura/etal:2006}.}
\label{fig:fig9}
\end{figure}

To understand this, one must notice that two boundary steps of a
hexagon, which meet at one corner, are geometrically inequivalent
relative to the substrate. These are often designated as A and B
steps, and the distinction between both becomes clear from
Fig.~\ref{fig:fig9}(a). If diffusion properties with respect to these
different step types were the same, islands would assume a hexagonal
shape.  Differences in the diffusion properties originate primarily
from three sources: (i) Stronger binding of adatoms to, say, A steps.
Attaching adatoms then have a tendency to enrich at A steps
\cite{Jacobsen/etal:1996}. (ii) Same binding energy, but faster
diffusion of adatoms along A, caused by a lower energy barrier
\cite{Michely/etal:1993}. This leads to a faster nucleation of new
atomic rows at A steps \cite{Michely/Krug:2004}. (iii) Asymmetric
corner diffusion, which means that a one-fold coordinated adatom at a
corner site goes preferentially to, say, A steps. Also this effect
leads to an enrichment of atoms at A steps. In all these cases,
further atoms are thus accumulating faster at one step type, chosen as
the A step here, and as a consequence the B steps grow at the expense
of A steps. Eventually triangular islands with prevailing B steps
form.

At low temperatures, the strong binding of atoms favors an aggregation
of hit-and-stick type, which leads to fractal-dendritic island shapes.
The overall symmetry and ramification of these structures depend again
on corner and edge diffusion properties. A demonstration of edge and
corner diffusion effects was given by \textcite{Ogura/etal:2006}, see
Fig.~\ref{fig:fig9}(b). For the islands shown in the upper panel of
this figure, corner diffusion is asymmetric with activation energies
$E_{\rm cB}<E_{\rm cA}$ [cf.\ Fig.~\ref{fig:fig9}(a)] while the island
shapes in the lower panel refer to a situation of symmetric corner
diffusion. From left to right the barrier $E_{\rm edge}$ [cf.\
Fig.~\ref{fig:fig9}(a)] for edge diffusion (along both A and B steps)
is lowered in both panels. With decreasing $E_{\rm edge}$, attaching
atoms can diffuse over longer distances along the edges and find more
favorable binding sites with higher coordination. As a consequence,
the structures become less ramified and the side arms thicken with
decreasing $E_{\rm edge}$. In the sequence of structures a skeleton
dendrite appears, which has triangular/hexagonal symmetry for
asymmetric/symmetric corner diffusion before eventually the compact
triangular/hexagonal shape is formed at even lower $E_{\rm edge}$.
Note that an overall triangular/hexagonal symmetry is also visible for
the island shapes simulated with large $E_{\rm edge}$.

It is also possible to obtain compact island shapes with curved edges,
if corner diffusion is suppressed \cite{Brune:1998}. By comparison
with measurements, features as the thickness of side branches or the
degree of ramification and further details of compact island shapes
can be used to identify kinetic parameters in specific models for
local diffusion at island boundaries \cite{Michely/Krug:2004}.

\subsection{Second-layer induced morphologies}
\label{sec:second_layer_induced_morphologies}

The second layer occupation can have a significant influence on the
shape and morphology of islands in the first layer. For example, the A
and B steps on (111) surfaces give rise to different step edge
barriers in the second layer. The downward fluxes across A and B steps
hence differ, which modifies the in-plane aspect ratio for island
shapes \cite{Li/etal:2008}.

\begin{figure}[t!]
\centering
\includegraphics[width=0.49\textwidth,clip=,]{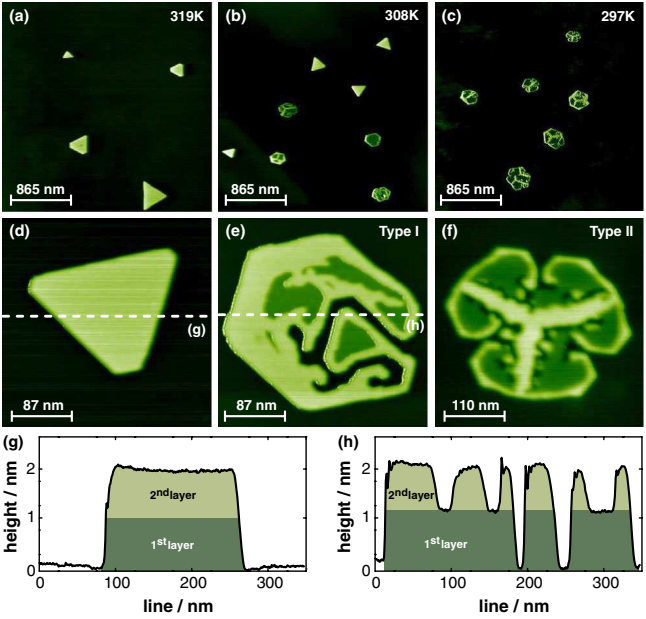}
\caption{(Color online) (a)-(c): AFM images of C$_{60}$ islands on
  CaF$_2$(111) at three different growth temperatures. (d)-(f):
  Magnified images of single islands: a compact triangle (d), and
  hexagonal islands with morphologies I (e) and II (f).  (g), (h):
  Height profiles along line scans shown in (d),(e).
  From \textcite{Koerner/etal:2011}.}
\label{fig:fig10}
\end{figure}

Quite unexpected complex morphologies can arise due to upward
transitions from the first to the second layer on weakly interacting
substrates. Such morphologies were recently found for fullerene
(C$_{60}$ molecules) adsorbed on ionic surfaces \cite{Burke/etal:2007,
  Burke/etal:2009, Loske/etal:2010}. Figure~\ref{fig:fig10} shows AFM
images of the self-assembly of C$_{60}$ after deposition on
CaF$_2$(111). At high temperatures [Fig.~\ref{fig:fig10}(a) and(d)]
triangular islands form that are two monolayers high
[Fig.~\ref{fig:fig10}(g)], while at lower temperatures islands with an
overall hexagonal shape emerge [Fig.~\ref{fig:fig10}(b),(c),(e), and
(f)] that have a base of one monolayer [Fig.~\ref{fig:fig10}(h)] and
exhibit a complicated structure with double layer rims at the island
edges and channels directed towards the interior of the islands. These
low-temperature morphologies are very different from the
fractal-dendritic island shapes found in metal epitaxy.

\begin{figure}[b!]
\centering
\includegraphics[width=0.28\textwidth,clip=,]{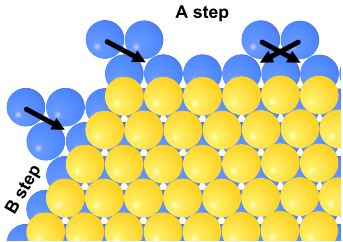}
\caption{(Color online) Sketch of elementary upward jump processes
  facilitating the growth of a double-layer triangle.  The processes
  at the A step and the corner between the A and B step involve only
  two C$_{60}$ and are, therefore, more likely than the process at the
  B step involving three C$_{60}$ molecules.  From
  \textcite{Koerner/etal:2011}.}
\label{fig:fig11}
\end{figure}

The emergence of these morphologies can be understood from a mechanism
of facilitated dewetting \cite{Koerner/etal:2011}. Facilitated
dewetting means that a C$_{60}$ molecule on an edge site in the second
layer lowers the energy barrier for an upward transition of another
molecule to a neighboring edge site in the second layer. At high
temperatures a first upward transition over a bare dewetting barrier
typically occurs early during growth of an island, when it consists of
only a few C$_{60}$ molecules. C$_{60}$ molecules in the second layer
subsequently lower the energy barrier for further upward transitions
to neighboring sites, and as a consequence islands grow as
double-layers. That these double-layer islands evolve into triangular
rather than hexagonal shapes has its origin, similar as for island
growth on (111) surfaces in metal epitaxy (cf.\
Sec.~\ref{sec:Anisotropy_in_corner_diffusion}), in a symmetry breaking
effect associated with A and B steps. The distinction between these
steps has no meaning here for mono-layer islands because of the large
diameter of the C$_{60}$ molecules compared to the lattice constant of
the CaF$_2$(111) substrate. The two types of steps can be
distinguished, however, in the second layer. As illustrated in
Fig.~\ref{fig:fig11}, upward transitions of C$_{60}$ at A steps are
more frequent than at B steps, because they require only two rather
than three C$_{60}$ molecules in the first layer. Accordingly,
triangular shapes with prevailing B steps result.

At low temperatures upward transition of C$_{60}$ over the bare
dewetting barrier become rare events and large islands with overall
hexagonal shape form. For these large islands, deposition on top of
islands is the dominating process of second layer occupation with
nucleation kinetics as described in Sec.~\ref{sec:second_layer} (for
$i=1$ and large step edge barriers). An island initially one layer
high can evolve into two different types of morphologies I and II,
depending on whether the first stable cluster of C$_{60}$ in the
second layer is nucleated close to an edge or close to the center of
the island.

Growth of a stable second layer cluster close to the island boundary
soon leads to a contact with an edge, which triggers the process of
facilitated dewetting. Starting form the point of contact,
double-layer rims are growing along the island edges due to
facilitated upward transitions of C$_{60}$. During the period of
C$_{60}$ deposition, these rims do not succeed to surround the island,
because the island area extends faster by newly attaching C$_{60}$ to
the rim-free edges than to the edges with rim, where facilitated
dewetting leads to a broadening of the rim. In the post-deposition
regime, however, the ends of the rim grow further by facilitated
upward transitions of C$_{60}$ that stem from the rim-free edges and
diffuse along the island edges. If the two ends of the rim eventually
approach each other along the same edge, a funnel starts to form and
further growth of this funnel leads to a trench extending towards the
interior of the island together with the rim. This leads to morphology
I with a typical example shown in Fig.\ref{fig:fig10}(e).

If a stable second layer cluster nucleates close to the island center,
a dendritic-skeletal cluster shape evolves, which resembles the
simulated morphology in the second last structure (from left to right)
of the upper panel of Fig.~\ref{fig:fig9}(b) in
Sec.~\ref{sec:Anisotropy_in_corner_diffusion}. In the post-deposition
period the skeletal dendrite grows very slowly by rare upward
transitions of C$_{60}$ over the bare dewetting barrier until one of
the three finger tips reaches an island edge. At this moment
facilitated dewetting transitions set in and a rim starts to grow
along both sides of the finger terminus at the island edge. The
formation of the rim is accompanied by a loss of C$_{60}$ molecules at
the rim-free edges causing one of the two other finger tips of the
dendrite to reach an island edge. A rim then starts to grow also from
this finger terminus and thereafter also the third finger tip reaches
the island edge with subsequent rim formation.  Eventually the three
growing rims move towards each other close to edge points located half
way between the tip termini, where funnels form and subsequently
trenches grow towards the island interior. As a result, morphology II
with an approximate threefold symmetry is obtained with a typical
example shown in Fig.~\ref{fig:fig10}(f).

All these complicated structures found in experiment were successfully
modeled by a kinetic growth model based on the mechanism of
facilitated dewetting \cite{Koerner/etal:2011}.

\subsection{Segregation and ordering effects}
\label{subsec:A3B}
Questions concerning the inner structure of clusters are most relevant
for 3d binary systems, produced by co-deposition of different atomic
species. Compositional fluctuations in these systems are characterized
primarily in terms of atomic short- or long-range order in the
cluster's interior, and in terms of surface segregation, i.~e.\ in the
enrichment of one atomic species at the cluster surface.  At
equilibrium these two features generally compete with each other
\cite{Polak/Rubinovich:2000}. This can be understood intuitively
because strong ordering interactions inside the ``bulk'' will enforce
atomic order up to the surface and hence impede surface segregation.
Conversely, surface interactions favoring segregation will suppress
ordering tendencies, at least in the near-surface region.

For cluster growth outside equilibrium, these arguments remain
qualitatively valid, but the degree of ordering and segregation
diminishes. Although atomic diffusion, necessary for equilibration, is
often frozen in the bulk, some remanent bulk order can develop, as a
result of previous diffusion steps of surface atoms before being
burried by the external flux.  So the question arises: What kind of
metastable compositional fluctuations are generated below the advancing
surface of a 3d nanocluster under the condition of active surface but
frozen bulk kinetics?

\begin{figure}[t!]
\centering
\includegraphics[width=0.45\textwidth,clip=,]{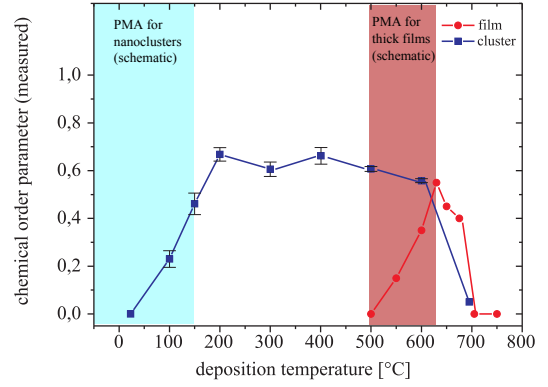}
\caption{(Color online) Chemical order for CoPt$_{3}$-clusters (grains) grown on
 WSe$_2$(0001) and CoPt$_{3}$-films grown on Pt(111)
 as a function of the growth temperature. $L1_{2}$-type
 ordering sets in below the bulk disorder/order transition
 temperature $T\simeq686^{\rm o}{\rm C}$, and vanishes at lower
 temperatures when the bulk kinetics becomes frozen.
 For clusters, the vanishing of L1$_2$-type ordering occurs at a
 lower temperature because atomic rearrangements at the advancing
 surface allow some degree of chemical order in the interior
 to be built. The occurrence of PMA is schematically indicated
 for both clusters and films.
 Adapted from~\textcite{Maier/etal:2002}.}
\label{fig:fig12}
\end{figure}

Theoretically, the relationship between surface kinetics and emerging
frozen bulk structure has remained largely unexplored.  Some aspects
were recently studied with the help of an analytically solvable model
for 1d growth \cite{Einax/Dieterich:2008}. Quite obviously, remanent
bulk order depends on the ratio between the time scales for surface
diffusion and atomic deposition. Coming from high temperatures,
ordering will initially improve upon cooling, but near some blocking
temperature, where those two time scales match, it will pass a maximum
and drop down to zero as $T\to0$ (see Fig.~\ref{fig:fig12} for an
example). In this limit both bulk and surface kinetics get frozen,
leading to random compositions.

Particularly interesting is the structure of alloy clusters with
magnetic components. Special attention have received attempts to
generate perpendicular magnetic anisotropy (PMA), where the easy axis
of magnetization is perpendicular to the substrate plane. This
requires the magneto-crystalline perpendicular anisotropy to be
stronger than the shape anisotropy due to dipolar interactions, which
generally favors in-plane magnetization. PMA is well known to occur in
multilayer films \cite{Johnson/etal:1996} and has been exploited to
increase storage capacities in magnetic devices. Its occurrence in
nanoclusters was detected, for example, in FePt or CoPt alloy
clusters. Depending on the technique and on the conditions of growth,
the experiments suggest an anisotropic short-range
\cite{Liscio/etal:2010} or layer L1$_0$-type long-range order
\cite{Zeng/etal:2002, Perumal/etal:2008, Andersson/etal:2006,
  Moulas/etal:2008} to be associated with the PMA.

Magneto-crystalline anisotropies are caused by quantum-mechanical
effects, which lead to a preferential alignment of magnetic moments
along symmetry directions in the crystal lattice. Most important are
hybridization of $d$ electron states between neighboring atoms and a
strong spin-orbit coupling. A theoretical description
requires sophisticated \textit{ab initio} calculations for the
electronic structure, together with the Dirac equation to include
relativistic effects [see, e.g., \textcite{Sipr/etal:2010}].

To get an understanding of the connection between the compositional
structure of clusters and the occurrence of PMA, one can adopt a
simplified bond picture \cite{Neel:1954,Victora/MacLaren:1993}. For
an $AB$ binary alloy nanocluster with vacancies $V$ the
magneto-crystalline anisotropy energy is then expressed as
\begin{equation}
\label{eq:HA}
  H_{\mathcal A} = - \sum_{\langle i,j\rangle}\sum_{\alpha,\beta}
  \mathcal{A}_{\alpha\beta}
(\hat{\boldsymbol\mu}_i^\alpha\cdot
\hat{\boldsymbol\delta}_{ij})^2\,
m_i^\alpha m_j^\beta\,,
\end{equation}
where the sum runs over all cluster sites $i$ and their
nearest-neighbor sites $j$, connected by bond vectors
${\boldsymbol\delta}_{ij}$; $m_i^\alpha$ are occupation numbers
($m_i^\alpha=1$ if site $i$ is occupied by species $\alpha$ or zero
else, $\alpha=A,\,B,\,V$), and ${\boldsymbol\mu}_i^\alpha$ are the
magnetic moments (${\boldsymbol\mu}_i^V=0$);
$\hat{\boldsymbol\delta}_{ij}$ and $\hat{\boldsymbol\mu}_i^\alpha$
designate unit vectors. The parameters $\mathcal{A}_{\alpha\beta}$
quantify the magnetic anisotropy energies for moments
${\boldsymbol\mu}_i^\alpha$ associated with an occupation of
neighboring sites by $\beta$ species. They can be estimated from
experiments or determined from \textit{ab initio} calculations. Note
that Eq.~(\ref{eq:HA}) represents the lowest order term of an
expansion in powers of magnetic moments implied by symmetry arguments.

Using Eq.~(\ref{eq:HA}), the structural magnetic anisotropy energy of
a fully magnetized cluster is given by
$E_{\rm str}= H_{\mathcal A}\{\mbox{all}\;{\boldsymbol\mu}_i^\alpha\;
\mbox{in plane}\}-
H_{\mathcal A}\{\mbox{all}\;{\boldsymbol\mu}_i^\alpha\; \mbox{out of plane}\}$.
This can be reduced to calculating the numbers
$n_\parallel^{\alpha\beta}$ and $n_\perp^{\alpha\beta}$ of in-plane
and out-of-plane $\alpha-\beta$ bonds with direction parallel to the
substrate and with components perpendicular to it, respectively.

\begin{figure}[t!]
\centering
\includegraphics[width=0.48\textwidth,clip=,]{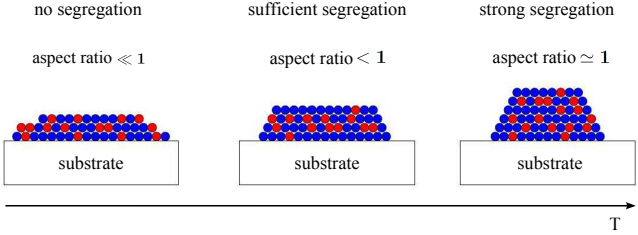}
\caption{(Color online) Sketch of changes in the structure and shape
  of binary alloy nanoclusters (e.g.\ CoPt$_3$) with temperature.
  Atoms of the two components are marked in red (e.g.\ Co) and blue
  (e.g.\ Pt), where the latter tend to segregate at the cluster
  surface. With increasing $T$ the fraction of side facets increases
  relative to the top facet and the surface segregation becomes
  stronger. The two competing effects for PMA lead to a temperature
  window of its occurrence.}
\label{fig:fig13}
\end{figure}

The procedure has been applied to study the occurrence of PMA for
chemically disordered CoPt$_{3}$-clusters, which were grown on a van
der Waals substrate \cite{Albrecht/etal:2001, Albrecht/etal:2002}.
These clusters are at low temperature in a metastable state, where the
formation of L$1_2$ type chemical order is kinetically suppressed. PMA
occurs below and near the onset of L$1_2$ ordering, see
Fig.~\ref{fig:fig12}. As observed for films, one can expect that this
PMA vanishes for even lower temperatures and hence occurs in a
temperature window. Using values reported for Co and Pt moments as
well as experimental results for Co-Pt multilayers and theoretical
results for the Co-vacuum interfaces, one can estimate that the
dominant contribution in Eq.~(\ref{eq:HA}) comes from Co-Pt bonds with
$\mathcal{A}_{\rm CoPt}\simeq250\,\mu{\rm eV}$, and accordingly
$E_{\rm str} \propto n_\perp^{\rm CoPt}-n_\parallel^{\rm CoPt}$
\cite{Heinrichs/etal:2007}. Taking into account the strong surface
segregation of Pt caused by its larger size compared to Co
\cite{Gauthier/etal:1992}, a mechanism for the PMA as depicted in
Fig.~\ref{fig:fig13} is conceivable. At low $T$, flat clusters with
extended top and small side facets occur and surface segregation is
kinetically suppressed. At high $T$ the side facets become
comparatively large. Depending on details of the interactions, an
intermediate temperature range can exist, where the cluster is still
fairly flat and the surface segregation is sufficiently strong. In
this case the contribution to the magnetic anisotropy energy coming
from the out-of-plane Co-Pt bonds at the top facet can be larger than
the contribution from the in-plane Co-Pt bonds at the side facets.
Accordingly, PMA is expected to occur.

Indeed, this mechanism for the occurrence of PMA could be corroborated
by KMC simulations of an $AB_3$ alloy with nearest neighbor
interactions $V_{AA}$, $V_{BB}$, and $V_{AB}$ adjusted to equilibrium
properties of CoPt$_{3}$
\cite{Heinrichs/etal:2006,Heinrichs/etal:2007}. The substrate was
modelled by a weak attractive substrate potential. Growth of the
clusters in time proceeds by co-deposition of Co and Pt atoms,
vacancy-assisted nearest neighbor hopping and by direct exchange
between unlike low-coordinated atoms on top of terraces or step edges.
Such direct exchange processes are often observed in heteroepitaxial
growth, and in particular were seen for Co deposited on Pt(111)
\cite{DeSantis/etal:2002,Gambardella/etal:2000}. As the model is
fully three-dimensional, interlayer diffusion and Ehrlich-Schwoebel
barriers are automatically taken into account.

\begin{figure}[t!]
\centering
\includegraphics[width=0.49\textwidth,clip=,]{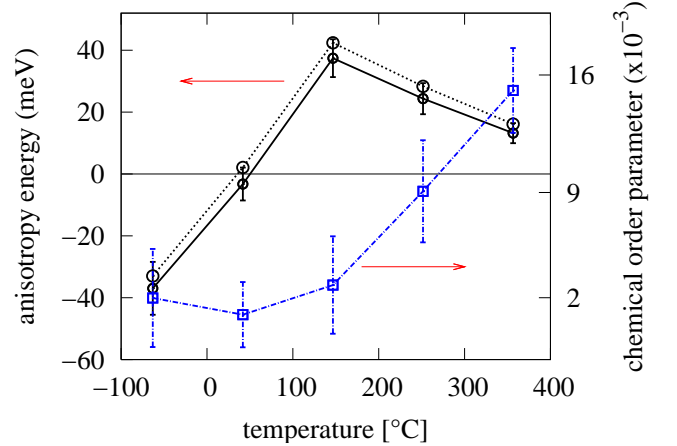}
\caption{(Color online) Magnetic anisotropy energies $E_{\rm tot}$
  (solid line) $E_{\rm str}$ (dotted line), and the chemical order parameter
  (dashed-dotted line) versus temperature of simulated CoPt$_3$ nanoclusters with
  $1000$ atoms for strong surface segregation of Pt
  ($V_{\rm PtPt}-V_{\rm CoCo}=180$~meV) and high exchange rates of Co
  and Pt at the cluster surface. Adapted from~\textcite{Heinrichs/etal:2007}.}
\label{fig:fig14}
\end{figure}

Figure~\ref{fig:fig14} shows model results for the magnetic anisotropy
energy, both for its structural contribution $E_{\rm str}$ and the
total energy $E_{\rm tot}=E_{\rm str}+E_{\rm dip}$ obtained by adding
the (negative) dipolar contribution. The maximum $E_{\rm
  str}\simeq40$~meV at $T\simeq145^{\rm o}\,{\rm C}$ was shown to
originate from the interplay of Pt surface segregation, facilitated by
direct exchange processes, with $T$-dependent cluster shapes. It hence
reflects the mechanism sketched in Fig.~\ref{fig:fig13} so that PMA
indeed is a surface effect.

Simulations with varying cluster size $N$ show that $E_{\rm str}
\propto N^{2/3}$, so that PMA indeed is a surface effect. The
$N$-dependence of the total anisotropy energy therefore is given by
\begin{equation}
\label{eq:Etot}
E_{\rm tot} = E_{\rm str} + E_{\rm dip} =
K_{\rm str} N^{2/3} - K_{\rm dip} N
\end{equation}
with anisotropy constants $K_{\rm str}$ and $ K_{\rm dip}$.
Equation~(\ref{eq:Etot}) was shown to represent the KMC data very
well. It predicts an optimal mean cluster size, where PMA is
strongest, and a critical mean cluster size, above which the
magnetization switches to an in-plane orientation (when $E_{\rm
  tot}<0$). These predictions have yet to be confirmed in
experiments. For cluster sizes $N \simeq 1200$, as used in the
measurements by \textcite{Albrecht/etal:2001}, $E_{\rm tot}$
calculated from the model has the right order of magnitude when
compared with the experimental one. Also shown in Fig.~\ref{fig:fig14}
is the structural order parameter for L1$_2$ ordering, which
quantifies the preferential occupation of one of the four simple cubic
sublattices of the fcc lattice by Co atoms. While L1$_2$ ordering
remains kinetically suppressed up to room temperature, it becomes
significant only at higher temperatures, where PMA has already passed
its maximum. These trends obviously agree with the experimental
behavior displayed in Fig.~\ref{fig:fig12}.

Let us remark that measurements of the local order in CoPt$_3$
nanostructures on a van der Waals substrate confirmed preferential
Co-Pt/Co-Co binding out-of-plane/in-plane as the origin of PMA at room
temperature \cite{Liscio/etal:2010}, but in addition uncovered
structural details which go beyond the above model. We further note
that \textcite{Maranville/etal:2006}, using simulations on a more
microscopic level, were able to interpret PMA in ultrathin continuous
CoPt$_{3}$ films occurring at elevated temperatures in terms of Co
segregation along step edges.

In this Section our main focus was on examples of unexpected cluster
shape and structure formation, as discovered in recent experiments.
Specific mechanisms were identified that can explain these
experiments. For fullerenes (C$_{60}$) adsorbed on a weakly
interacting substrate, see
Sec.~\ref{sec:second_layer_induced_morphologies}, a distinction
between A- and B-steps and anisotropic growth arises through second
layer occupation. This is in contrast to
Sec.~\ref{sec:Anisotropy_in_corner_diffusion}, where the distinction
between A- and B-steps is due to the substrate. In turn, a new
mechanism of second-layer facilitated dewetting was proposed, which
explains the triangular to hexagonal shape transition of fullerene
islands, followed in the post-deposition regime by the evolution of
complex morphologies.

To unreveal the possible origins of PMA in nanoclusters, CoPt$_3$
nanocluster growth on a van der Waals substrate was considered in
Sec.~\ref{subsec:A3B}. It turned out that PMA occurs as a result of
an anisotropic atomic short range order in the cluster. Important
features in the clusters' atomic structure and their magnetic
properties were interpreted by a mechanism based on active surface but
frozen bulk kinetics and a concomitant competition between cluster
shapes and Pt surface segregation.

\section{Concluding remarks}
\label{sec:conclude}
Cluster growth on surfaces is a field that connects fundamental
studies of non-equilibrium phenomena with questions related to the
development of nanomaterials of practical use. Modern topics such as
growth of organic molecules and of nanoalloys with functional
properties make it necessary to reanalyze specific questions which are
central to this field. In this Colloquium we summarized basic concepts
of surface growth kinetics and showed for a number of examples how
these can be extended and further developed to tackle open problems of
current interest.

An important basis for describing cluster growth on surfaces is laid
by the rate equation approach. Extending this approach to binary
alloys, or, more generally, to multi-component adsorbates provides an
accurate framework for the analysis of future experiments on the
submonolayer kinetics driven by codeposition of two (or more) atomic
species. We hope that this framework, presented in
Sec.~\ref{sec:islands}, will stimulate experimental tests of our
findings and eventually will help to control nanoalloy surface growth.

Even for one-component metallic growth the exact behavior of the ISD
is still not known, not even in the $D/F \rightarrow \infty$ limit.
Section~\ref{sec:size_distribution} provides evidence from simulations
that rate equations based on ``correct'' capture numbers do have
predictive power for the ISDs. With respect to extensions to
multi-component systems, this feature is expected to remain valid.
However, for predicting ISDs from the rate equations, no analytical
theory of sufficient accuracy exists so far for the capture numbers.
Promising approaches for resolving this problem are theories for joint
probabilities of island size and capture area. Further developments of
such theories may provide a route also to account for coverage
dependencies of scaled ISDs in the $D/F\to\infty$ limit.

Nucleation of stable islands in confined geometries can be dominated
by rare fluctuations with the consequence that mean-field type
descriptions fail. This fact is particularly important for the problem
of second layer nucleation when the size of the critical nucleus is
one or two (Sec.~\ref{sec:second_layer}). Stochastic methods developed
for treating rare fluctuations proved applicable also to organic thin
film growth, when bending energies are involved in the passing of step
edges, and they were useful for making progress in other contexts such
as chemical reaction kinetics. There is yet more to be explored. For
example, rare fluctuations should play an important role also for
island nucleation on reconstructed surfaces, and different
Ehrlich-Schwoebel barriers for different type of atoms should have a
relevant influence on island shapes, similar to different barriers
associated with A and B steps on (111)-surfaces
\cite{Evans/etal:2006}.

Alloy cluster formation under non-equilibrium growth conditions allows
one to generate new atomic configurations that, while not relaxed in
thermal equilibrium, are nevertheless long-living due to frozen
kinetics. This is of particular interest when materials with new
functional properties can be created. An example is the occurrence of
PMA in alloy nanoclusters with components carrying magnetic moments
(Sec.~\ref{sec:3-D}). It was predicted that PMA can be enhanced when
clusters are grown in an external perpendicular magnetic field
\cite{Einax/etal:2007b,Einax/etal:2007c}, but this has not yet been
confirmed by experiment. For AB$_3$ alloys, which show L1$_2$-type
ordering at equilibrium, this enhancement is expected to be small. It
may become significant, however, for AB alloys exhibiting L1$_0$-type
ordering, as, for example, CoPt or FePt \cite{Lyubina/etal:2011}. At
equilibrium, these alloys display a transition to a layer structure
with alternating Co(Fe)- and Pt-rich layers. L1$_0$-type ordering
implies that the magnetic anisotropy becomes a bulk property, much
larger than that in CoPt$_3$. Since the appearance of the L1$_0$
phase requires relatively high annealing temperatures
\cite{Perumal/etal:2008, Markarov/etal:2009}, the temperature range
below the onset of long-range chemical order is of practical interest
as well. Regarding growth experiments under such conditions, these
materials seem to be promising candidates for detecting a substantial
magnetic field-induced enhancement of PMA. More experimental and
theoretical studies of the inner structure of metastable alloy
clusters grown by atomic deposition appear to be necessary in order to
exploit the full potential of these systems in materials science.

For organic thin film growth, the established concepts of submonolayer
growth kinetics should be revisited in order to incorporate
inter-molecular interaction effects such as $\pi$-stacking and
hydrogen bonding, which are absent in metal and semiconductor
adsorbates. In addition, new degrees of freedom, like rotation and
bending of molecules, need to be considered. Because of orientational
constraints for bonding and strong incommensurabilities of molecular
sizes (or of sizes of molecular subgroups) with substrate lattice
constants, critical nuclei can be composed of quite a large number of
molecules. These larger sizes of critical nuclei may be considered as
an intermediate case \cite{Schwarz/etal:2012} between the small
critical nuclei in metal growth and the large critical nuclei in
three-dimensional crystallite formation from solution. Theoretical
descriptions of the consequences of these and other features of
molecular adsorbates for the growth kinetics are only at the
beginning.  Novel island morphologies can emerge already due to weak
substrate-molecule interactions and an associated upward transport of
molecules from the first to the second layer. An unsettled question is
to what degree second-layer facilitated dewetting transitions, as
found for fullerenes on insulting surfaces, are a rather generic
mechanism influencing shapes of molecular clusters.

\section*{Acknowledgments}
We have greatly benefited from numerous discussions with our
colleagues and collaborators M.~Albrecht, S.~Heinrichs, A.~Majhofer,
M.~K\"orner, A.~K\"uhnle, M.~\mbox{Reichling}, J.~Rottler, and G.~Schatz.

\bibliographystyle{apsrmp4-1}
\bibliography{einax-etal-lit} 

\end{document}